\documentclass[aps,twocolumn,prd,superscriptaddress,showpacs,preprintnumbers,
amsmath,amssymb,nofootinbib, 10pt]{revtex4-1}

\usepackage{color}
\usepackage{graphicx}
\usepackage{amsmath, amssymb}
\usepackage{amsfonts}
\usepackage{dcolumn}
\usepackage{subfigure}
\usepackage{hyperref}
\usepackage{dsfont}

\newcommand{\Tr}{\ensuremath{\operatorname{Tr}}}
\def\eq#1{(\ref{#1})}
\def\Eq#1{Eq.~(\ref{#1})}
\def\fig#1{Fig.~\ref{#1}}
\def\sec#1{Sec.~\ref{#1}}
\def\id{\mathds{1}}

\def\lA0{{\langle A_0 \rangle}}
\def\bA0{{\bar{A}_0}}
\def\lLA{{\langle L[A_0] \rangle}}
\def\lL{{\langle L \rangle}}
\def\lLc{{\langle L^\dagger \rangle}}
\def\0#1#2{\frac{#1}{#2}}

\graphicspath{{./figures/}{./}}

\begin{document}

\title{Confinement order parameters and fluctuations}

\author{Tina Katharina Herbst}
\affiliation{Institut f\"ur Theoretische Physik,
Universit\"at Heidelberg, 
Philosophenweg 16, D-69120 Heidelberg, Germany}

\author{Jan Luecker}
\affiliation{Institut f\"ur Theoretische Physik, Universit\"at Heidelberg, 
Philosophenweg 16, D-69120 Heidelberg, Germany}
\affiliation{Institut f\"ur Theoretische Physik, Goethe-Universit\"at Frankfurt, 
Max-von-Laue-Stra{\ss}e 1 , D-60438 Frankfurt/Main, Germany}

\author{Jan M. Pawlowski}
\affiliation{Institut f\"ur Theoretische Physik, Universit\"at Heidelberg, 
Philosophenweg 16, D-69120 Heidelberg, Germany}
\affiliation{ExtreMe Matter Institute EMMI, GSI Helmholtzzentrum f\"ur 
Schwerionenforschung mbH, Planckstra{\ss}e 1, D-64291 Darmstadt, Germany}


%%%%%%%%%%%%%%%%%%%%%%%%%%%%%%%%%%%%%%%%%%%%%%%%%%%%%%%%%%%%%%%%%%%%%%%%%%%%%%%%%%%%%%%%%%%%%%%%%%%%
\begin{abstract}
  We study order parameters for the confinement-deconfinement phase
  transition related to the Polyakov-loop variable. The functional
  renormalisation group is used to compute these order parameters in a
  unified, non-perturbative continuum approach. Our result for the
  expectation value of the traced Polyakov loop agrees quantitatively
  with the lattice result. Furthermore, we discuss how this order
  parameter differs from the standard continuum Polyakov loop. For
  temperatures close to the phase transition temperature there are
  significant deviations. We argue that these deviations are of
  crucial importance for QCD effective models, which usually
  implicitly rely on a Gau\ss ian approximation neglecting this
  difference.
\end{abstract}

\maketitle

%%%%%%%%%%%%%%%%%%%%%%%%%%%%%%%%%%%%%%%%%%%%%%%%%%%%%%%%%%%%%%%%%%%%%%%%%%%%%%%%%%%%%%%%%%%%%%%%%%%%
%% Sec: Intro
%%%%%%%%%%%%%%%%%%%%%%%%%%%%%%%%%%%%%%%%%%%%%%%%%%%%%%%%%%%%%%%%%%%%%%%%%%%%%%%%%%%%%%%%%%%%%%%%%%%%
\section{Introduction}\label{sec:intro}

The physics of low energy QCD is mainly governed by two phenomena:
spontaneous chiral symmetry breaking and confinement. Order parameters
for the former are local observables that are not invariant under
chiral transformations, most prominently the chiral condensate. In
turn, a smoking gun for confinement is the linear rise of the static
quark-potential at infinite distance in pure $SU(N)$ Yang-Mills
theory. The underlying symmetry is the $Z_N$-center symmetry, which is
only accessible via non-local observables such as the traced Polyakov
loop. In QCD at the physical point, none of these symmetries is fully
realised and the respective phase transitions turn into crossovers:
chiral symmetry is explicitly violated by the current quark masses and
the chiral anomaly, and center symmetry is violated explicitly by
dynamical quarks in the fundamental representation. Nonetheless, the
respective order parameters are interesting observables in full QCD as
they give access to the amount of explicit symmetry breaking, and
hence the proximity of QCD at finite temperature and density to a
potential critical end-point in the phase diagram.

The confinement-deconfinement phase transition, or more precisely the
transition from the quark into the hadronic phase, is also accessible
with baryonic observables such as baryonic fluctuations. The latter
are directly accessible within experiments and are hence especially
interesting. In this context, observables based on the Polyakov-loop
variable are of specific interest as the related gluonic backgrounds
play an important r$\hat{\rm o}$le in the computation of the baryonic
fluctuations, see \cite{Fu:2015naa}.

More generally, in functional approaches to QCD the temporal gluonic
background $\langle A_0\rangle$ relates to the Polyakov loop. It
provides a natural expansion point for systematic expansions of
first-principle QCD, as well as for QCD-enhanced low-energy effective
models, see e.\,g. \cite{Braun:2009gm, Pawlowski:2010ht, Haas:2013qwp,
  Herbst:2013ufa, Mitter:2014wpa, Braun:2014ata} and the recent
survey \cite{Pawlowski:2014aha}.  This approach has been
established in
\cite{Braun:2007bx,Marhauser:2008fz,Braun:2009gm,Braun:2010cy,Fister:2013bh}. 
There it has been worked out how to define and compute the gauge
invariant non-perturbative glue potential $V[A_0]$ in Yang-Mills
theory and QCD within the functional renormalisation group, as well as
general functional methods. It has also been shown that the
expectation value $\langle A_0\rangle$ of the gluonic background,
defined as the minimum of the glue potential $V[A_0]$, serves as a
gauge-invariant order parameter for the confinement-deconfinement
phase transition. It is indeed directly linked to the gauge-invariant
eigenvalues of the untraced Polyakov loop. Moreover, the setting
enabled us to link the confinement-deconfinement phase transition
algebraically to a mass gap in the gluon propagator,
\cite{Braun:2007bx, Fister:2013bh}. In the latter work
\cite{Fister:2013bh}, the framework has been extended to general
functional approaches including the Dyson-Schwinger equations and the
2PI formalism, for applications to QCD see
\cite{Fischer:2013eca,Fischer:2014vxa,Fischer:2014ata}. In
\cite{Reinhardt:2012qe,Reinhardt:2013iia,Heffner:2015zna} the approach
has been extended to the Coulomb gauge in the Hamiltonian
formulation. It has also been picked up in more phenomenological
applications to QCD, see e.\,g. \cite{Kondo:2010ts, Fukushima:2012qa,
  Reinosa:2014ooa, Reinosa:2014zta, Kondo:2015noa}.

In the present work we study the relation between the expectation
value of the traced Polyakov loop and the expectation value $\langle
A_0\rangle$, the temporal gluonic background of the theory, including
the gauge invariant definition of the latter. In
Secs.~\ref{sec:OrderParams} and \ref{sec:OrderParamsgauge} of this
work we show how these order parameters for the
confinement-deconfinement phase transition are derived from the
Polyakov loop, and discuss their properties and the underlying
symmetries. Subsequently we set up the functional renormalisation
group (FRG) as a framework to study these objects in a unified
approach, see \sec{sec:PLFlows}.  Notably, we present the first
non-perturbative continuum calculation of the expectation value of the
traced Polyakov loop, which is a standard observable on the
lattice. We discuss the renormalisation of this object in
\sec{sec:Renormalization} and demonstrate the quantitative agreement
with the lattice results in \sec{sec:Comparison}.

\section{Order parameters and the Polyakov loop}
\label{sec:OrderParams}
In the present work we compute and compare different order parameters
for the confinement-deconfinement phase transition that can be derived
from the Polyakov loop. The properties of the Polyakov loop itself and
related constructions of order parameters have been discussed at
length in the literature. This includes the discussion of simple
representations in physical gauges such as the Polyakov gauge, maximal
Abelian gauges and axial gauges The latter gauges, and in particular
the Polyakov gauge, also facilitate the access to topological
excitations or defects and their r$\hat{\rm o}$le for the
confinement-deconfinement phase transition, see e.g.\
\cite{Ford:1998bt,vanBaal:2000zc} and references therein. Here we
briefly review some important properties.

The Polyakov loop, the Wilson loop in the temporal direction, is
defined as
\begin{equation} 
  \label{eq:Poloop} 
  P(\vec x)= {\cal P} \exp\left\{ i g \int_0^{\beta} dx_0\, A_0(x_0,\vec x)
  \right\} \,,\quad{\rm with}\quad \beta=\0{1}{T}\,,
\end{equation}
where $\cal P$ denotes path ordering, and $g$ the gauge coupling. Due
to the periodicity in the temporal direction, $x_0 \to x_0+\beta$,
gauge fields have to be periodic up to gauge transformations, the
temporal transition functions $t_0(x)$. The latter can be chosen
periodic, $t_0(x_0+\beta,\vec x)= t_0(x_0,\vec x)$. Under general
gauge transformations $U(x_0,\vec x)$, the Polyakov loop transforms as
\begin{equation}
  P(\vec x)\to U^{-1}(0,\vec x)P(\vec x) U(\beta,\vec x)\,.
\end{equation}
Notably, the traced Polyakov loop, $\Tr\, P(\vec x)$, is only
invariant under periodic gauge transformations with $U(x_0+\beta,\vec
x)= U(x_0,\vec x)$, that also guarantee the periodicity of the
transition function $t_0$. Under these periodic gauge transformations
\eq{eq:Poloop} transforms as a tensor 
\begin{equation}
  \label{eq:Utrafo}
  P(\vec x)\to U^{-1}(\vec x)P(\vec x) U(\vec x)\,,\ \ \text{with}\ \ 
  U(\vec x)= U(0,\vec x)\,, 
\end{equation}
and the traced Polyakov loop is invariant. In turn, under non-periodic
gauge transformations the periodicity of the temporal transition
functions changes and $P(\vec x)$ has to be augmented by transition
functions in order to provide a tensor under gauge
transformations. This generalisation is necessary e.\,g.  if a
temporal axial gauge is considered with $A_0=0$. Clearly $ P(\vec
x)=\id$ in such a gauge and carries no physics information. Then, the
information about the phase transition is comprised entirely in the
transition functions, see e.\,g.~\cite{Ford:1998bt} for more details.

Here, however, we resort to the periodic setting. Then one well-known
standard order parameter is given by the expectation value of the
trace of the Polyakov loop. The normalised trace reads
\begin{equation}
   \label{eq:tracedPL}
  L[A_0] = \frac{1}{N} \Tr_f P(\vec x)\,, 
\end{equation} 
which is gauge invariant under periodic gauge transformations. The
expectation value of \eq{eq:tracedPL}, $\lLA$, relates to the free
energy, $F_{q \bar q}$, of a static quark--anti-quark pair at infinite
distance,
\begin{equation}
  \label{eq:freen}
  \lLA \sim e^{-\012 \beta F_{q\bar q}}\,.
\end{equation}
Since the free energy of a quark--anti-quark pair is finite in the
deconfined phase and diverges in the confined phase, one can deduce
that $\lLA$ indeed is an order parameter for the
confinement-deconfinement transition. More precisely, it is an order
parameter for center symmetry breaking: transformations with $z\in Z_N$
transform the Polyakov loop into
\begin{equation} 
  \label{eq:ztrafo} 
  P(\vec x)\to z P(\vec x)\quad \Rightarrow \quad L[A_0]\to z L[A_0] \,,
\end{equation}
where $z$ in \eq{eq:ztrafo} is in the fundamental
representation. Hence, in the center-symmetric phase the expectation
value of $L$ has to vanish. In turn, the ground state in the
perturbative high-temperature phase is given by vanishing gauge fields
which break center symmetry maximally. Then we have $\lLA>0$, if we
single out the group direction that gives real and positive values for
the Polyakov loop. This can either be done by a small explicit
symmetry breaking in the effective Polyakov loop potential or by
appropriate temporal boundary conditions, both are common
procedures for computing the order parameters of spontaneous symmetry 
breaking.  We are finally led to
\begin{equation}
  \label{eq:exptrL}
  \lLA  \ \left\{\begin{array}{rcl} = 0 & \quad  {\rm for}\quad & T<T_c\\[2ex] 
      > 0 & \quad {\rm for}\quad & T>T_c\end{array}\right. \,.
\end{equation}
As the underlying symmetry is the discrete $Z_N$-symmetry, we expect a
second-order phase transition for $SU(2)$ and a first-order transition
for $SU(N>2)$. This has indeed been verified on the lattice as well as
in the continuum. Moreover, the temperature-dependence of the order
parameter shows rapid convergences towards the large-$N$ limit, for
results in the present continuum QCD setting and on the lattice see
e.\,g.~\cite{Braun:2010cy,Mykkanen:2012ri}. Despite this successful
classification of the confinement-deconfinement phase transition in
semi-simple Lie groups there are strong indications that it is the
number of degrees of freedom that triggers the order of the
confinement-deconfinement phase transition, rather than the center.
For example, for exceptional Lie groups one sees first-order phase
transitions, e.\,g.~\cite{Braun:2010cy}, while for $SO(3)$ a second
order phase transition is observed, e.\,g.~\cite{deForcrand:2002vs}.

\begin{figure*}
  \includegraphics[height=.26\textheight]{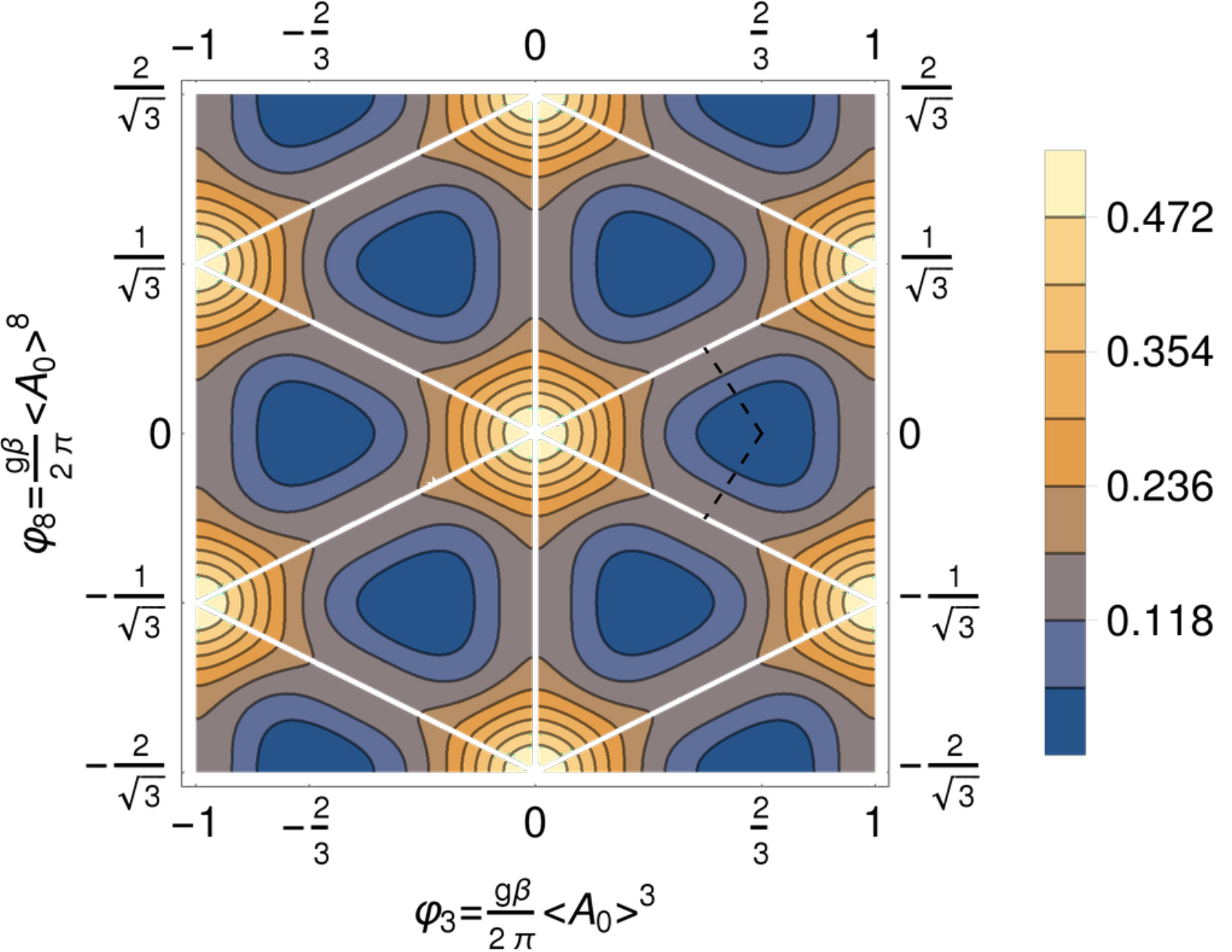}\hspace{20pt} 
  \includegraphics[height=.26\textheight]{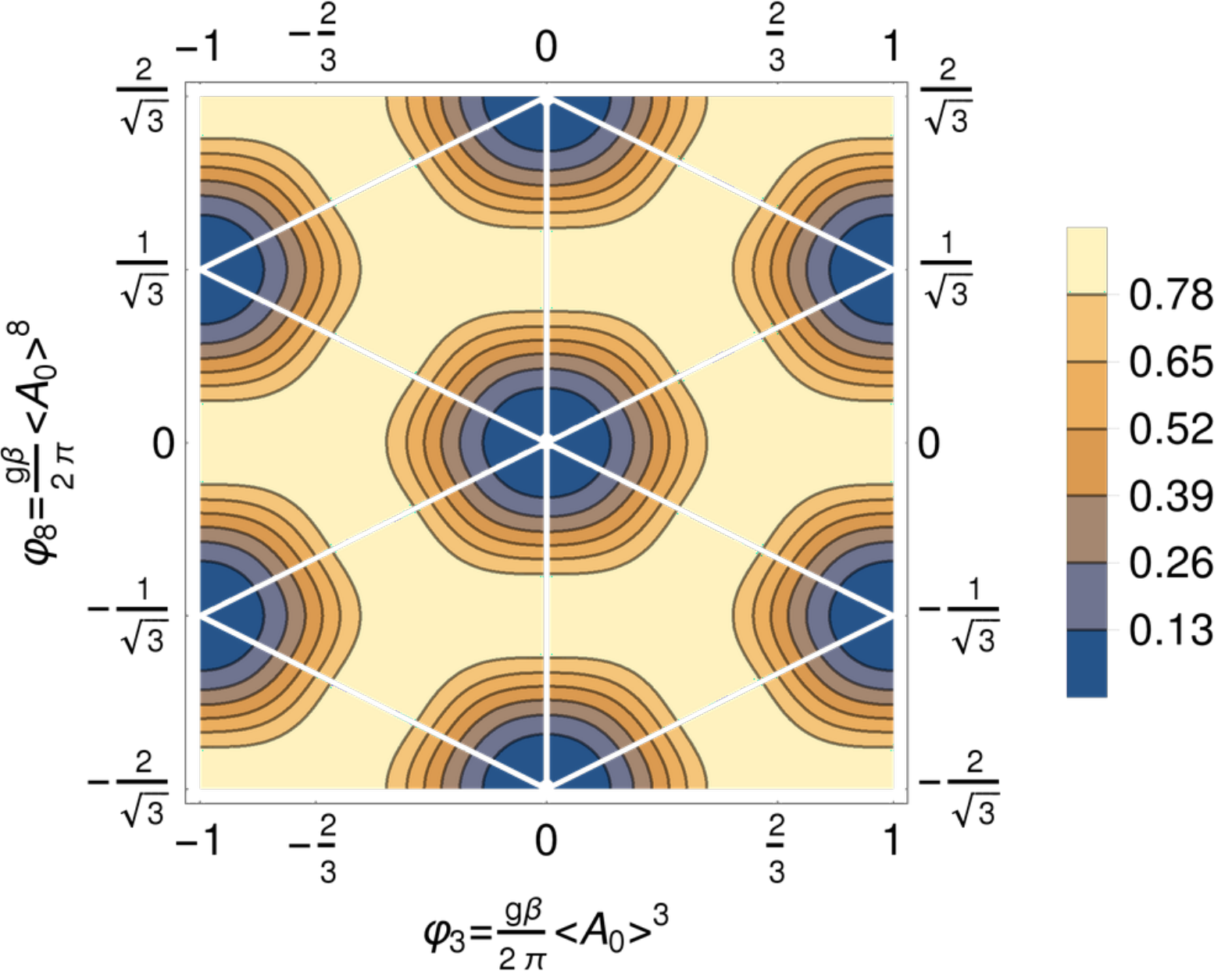}
  \caption{Equipotential lines of the glue potential,
    $V(\varphi_3,\varphi_8)$, in the confined phase (left,
    $T=236$~MeV) and in the deconfined phase (right,
    $T=384$~MeV). Lighter colours indicate higher potential values. The
    periodicity of the potential is clearly visible. The triangles are
    the Weyl chambers and Weyl reflections are the reflections about the
    edges of the Weyl chambers.  The centers of the Weyl chambers
    single out the center symmetric points of the Polyakov loop.}
  \label{fig:VglueContours}
\end{figure*}
The expectation value of the traced Polyakov loop \eq{eq:exptrL} is a
simple order parameter for the confinement-deconfinement phase
transition which is easily accessible on the lattice. In terms of the
gauge field, however, it requires the computation of infinite-order
correlation functions of the temporal gauge field. Hence, within
continuum computations based on a description in gauge fields this
asks for the definition of an order parameter which is more easily
accessible. To this end we note that the Polyakov loop \eq{eq:Poloop} 
can also be written as the exponential of an algebra-valued field $\varphi$,
\begin{equation}
  \label{eq:varphiP}
  P(\vec x)= e^{ 2 \pi i\, \varphi(\vec x)}\,,\quad {\rm with}\quad 
  \varphi(\vec x)\to U^{-1}(\vec x)\, \varphi(\vec x)\, U(\vec x)\,. 
\end{equation}
The transformation property of $\varphi$ in \eq{eq:varphiP} follows
straightforwardly from that of the Polyakov loop
\eq{eq:Utrafo}. Hence, $\varphi$ also transforms as a tensor under
gauge transformations. This entails that the eigenvalues $\nu_n$ of
$\varphi$ are gauge invariant, as gauge transformations are unitary
rotations that do not change the eigenvalues of a matrix. We utilise
this freedom and rotate $\varphi$ into the Cartan subalgebra. We
shall briefly discuss below that this amounts to taking the Polyakov
gauge. Then, the gauge-invariant eigenvalues are given by the relation
\begin{equation}
  \label{eq:nun}
 \varphi(\vec x) |\phi_n\rangle  = \nu_n(\vec x) |\phi_n \rangle \,,
\end{equation}
with eigenvectors $|\phi_n\rangle$ that span the Cartan
subalgebra. This allows us to define a constant background
\begin{equation}
  \label{eq:barvarphi}
  \bar\varphi= \sum_n \langle \nu_n \rangle |\phi_n 
  \rangle\langle  \phi_n |\,,
\end{equation} 
which carries the gauge invariant information about the eigenvalues of
the Polyakov loop, $\exp \{2 \pi i\, \nu_n\}$, and hence about the
confinement-deconfinement phase transition. Note that for the
expectation values $\langle \nu_n\rangle$ a preferred group direction has to
be singled out, similarly to the computation of the expectation value
of the traced Polyakov loop.

Indeed, $\bar\varphi$ itself is
an order parameter, which can be deduced from the transformation
properties of $\varphi$ under center transformations. Before this is
detailed, let us remark that in the Polyakov gauge there is a simple
relation between the algebra-valued field $\varphi$ and the gauge field. 
The Polyakov gauge reads
\begin{equation}
  \label{eq:polgauge} 
  A_0(x) = A_0^c(\vec x)\,,
\end{equation} 
where $A_0$ depends on the spatial coordinates $\vec x$ only and is
rotated into the Cartan subalgebra indicated by the superscript $c$ in
\eq{eq:polgauge}. Then we have
\begin{equation}
  \label{eq:varphiPolgauge} 
  2\pi\varphi(\vec x)= g\beta A_0^c(\vec x)\,. 
\end{equation}
The relation \eq{eq:varphiPolgauge} entails that the eigenvalues of $
g\beta A_0^c(\vec x)$ in Polyakov gauge are the gauge-invariant
eigenvalues of $\varphi$. Hence, despite working in a gauge-fixed
setting, we can directly extract gauge-invariant information from the
expectation value of the gauge field. With \eq{eq:varphiPolgauge} the
traced Polyakov loop \eq{eq:tracedPL} then takes the simple form
\begin{equation}
  \label{eq:LA0}
  L[A_0]=\01N \Tr_f e^{ig\beta A_0^c} =\01N \Tr_f e^{2\pi i \varphi} =:L(\varphi)\,.
\end{equation} 
Note also that \eq{eq:varphiPolgauge} holds for any constant gauge
field as constant fields can be rotated in the Cartan subalgebra. For these
fields in $SU(2)$ we have $\beta g A_0= 2 \pi \varphi_3\, \sigma^3/2$
with the generator $\sigma^3/2$ of the Cartan subalgebra. For $SU(3)$
we write
\begin{equation}
  \label{eq:varphi}
  \beta g A_0 = 2\pi\left(\varphi_3t^3 + \varphi_8t^8\right) =: 2\pi
  \varphi\,,
\end{equation} 
with $t^{3},\,t^8$ the generators of the Cartan subalgebra of
$SU(3)$. With \eq{eq:varphi} the traced Polyakov loop in $SU(3)$ 
Yang-Mills theory takes the form
\begin{equation}
  \label{eq:Lvarphi}
  L(\varphi_3,\varphi_8) = \013 \left( e^{- \0{2 \pi i \varphi_8}{\sqrt{3}} } +2\cos( 
  \pi\varphi_3)\, e^{\0{\pi i \varphi_8}{\sqrt{3}} } \right)\,.
\end{equation}
The above entails that the
corresponding effective potential $V(\varphi)$ simply is that of
constant temporal gauge fields $V[A_0]$. The expectation value
$\bar\varphi$ can be determined from the minimum of the effective
potential $V(\varphi)$. As for $\lLA$, we single out the minimum with
$L(\bar\varphi)\in \mathbb{R}$ and positive, leading to $\varphi_8=0$,
see \eq{eq:Lvarphi}.  Contour plots of the effective potential both in
the deconfined and in the confined phase are shown in
\fig{fig:VglueContours}, and details about its continuum definition
are provided in the next section. In the confined phase we
have $\bar\varphi_3=2/3$ and $L[\bar\varphi]=0$ while in the
deconfined phase we have $0\leq\bar\varphi_3<2/3$, $L[\bar\varphi]\neq
0$, with $\bar\varphi_8=0$ in both cases. This already indicates that
$\bar\varphi$ as well as $L[\bar\varphi]$ are order parameters for
center symmetry breaking.

Now we turn to the properties of $\varphi$ and the expectation value
$\bar\varphi$ under symmetry transformations, and in particular center
transformations. On the level of the algebra field $\varphi$, center
symmetry is realised as a shift
\begin{equation}
  \label{eq:thetaz} 
  \varphi\to \varphi+ \theta_z\,,\quad {\rm with} \quad 
  z=e^{2 \pi i \theta_z}\,.
\end{equation} 
The center transformation is a symmetry transformation of the
effective potential $V(\varphi)$, as can be seen from the contour
plots in \fig{fig:VglueContours}.  For $SU(2)$ the center elements are
given by $z=\id, -\id$. The corresponding algebra elements of $z$ are
$\theta_z = 0, \sigma^3/2$ respectively, with the Pauli
matrix $\sigma_3$. For $SU(3)$ the center elements $z$ 
and the corresponding algebra elements $\theta_z$ are given by 
\begin{equation}
  \label{eq:thetazSU3}
  z=\id\,,\ \id e^{\0{2 \pi}{3} i}\,, \ \id e^{\0{4 \pi}{3} i }\,,\qquad  \theta_z =
  0\,, \  \0{2}{\sqrt{3}}\, t^8 \,,\  t^3 - \0{1}{\sqrt{3}}\,t^8\,,
\end{equation}
respectively with the Cartan generators $t^3,t^8$ of $SU(3)$ being half of the
Gell-Mann matrices $\lambda^3,\lambda^8$. 

The transformation \eq{eq:thetaz} can be combined with a Weyl
reflection, i.e. a reflection about the edges of the Weyl chamber.
Weyl reflections are an isometry of the roots of the gauge group and
are generated by specific constant gauge transformations. Hence they
are a symmetry of the theory. In $SU(2)$ they are simply given by
$\varphi\to -\varphi$, while in $SU(3)$ the Weyl reflections can be
read off from \fig{fig:VglueContours}.  Clearly this is a symmetry of
the potential. Restricting ourselves to the fundamental chamber, they
map the chamber onto itself, and the centers of the Weyl chambers are
fixed-points of the combined symmetry transformations, see e.g.\
\cite{vanBaal:2000zc} for a more detailed discussion. Hence, in the
center-symmetric phase $\bar\varphi$ has to settle at these points. We
conclude that $\bar\varphi$ and the Polyakov loop $L[\bar\varphi]$ are indeed
order parameters for the confinement-deconfinement phase transition,
\begin{equation}
  \label{eq:barL}
  L[\bar\varphi]\  \left\{\begin{array}{rcl} = 0 & \quad  {\rm for}\quad & T<T_c\\[2ex] 
      > 0 & \quad {\rm for}\quad & T>T_c\end{array}\right. \,,
\end{equation}
where we have singled out the positive semi-definite minimum on the
real axis, i.e. $\bar\varphi^8=0$.

\section{Order parameters and gauge fields}\label{sec:OrderParamsgauge}
In the following we compute the order parameters, \eq{eq:exptrL} and
\eq{eq:barL}, within a non-perturbative approach to continuum
Yang-Mills theory formulated in terms of the gauge fields
$A_\mu$. Both order parameters can be expressed in terms of the
temporal gauge field $A_0$ and expectation values of correlations of
$A_0$. We have already discussed that the computation of $\lLA$ within
an expansion in the gauge field requires the computation of
infinite-order correlation functions of the gauge field. On the
lattice, on the other hand, it is straightforward to compute $\lLA$,
see e.g.\ \cite{Boyd:1996bx, Kaczmarek:2002mc, Zantow:2003uh,
  Gupta:2006qm, Gupta:2007ax, Dumitru:2003hp, Gavai:2010qd}.  In this
manner, the first-order nature of the phase transition in $SU(3)$ with
critical temperature $T_c/\sqrt{\sigma}=0.646$ in units of the string
tension $\sigma$ has been established. We shall see in
\sec{sec:flowPol} that within the functional renormalisation group the
task of computing $\lLA$ in the continuum is also tractable. It can be
reduced to solving a differential equation that is linear in the
scale-derivative of $L$ and its second $\varphi$-derivatives.

In turn, the computation of the order parameter \eq{eq:barL} only
requires the computation of the non-perturbative expectation value of
the constant temporal background gauge field in a background-field
approach, cf. \cite{Braun:2007bx, Fister:2013bh}, or in the Polyakov
gauge \cite{Marhauser:2008fz},
\begin{equation}
   \label{eq:contPL}
   L[\lA0] = \frac{1}{N}\Tr_f e^{i g\beta \lA0 }\,.
 \end{equation}
 The expectation value $\lA0$ is determined as the minimum of the
 non-perturbative effective glue background potential $V[\bar A_0]$ in
 the background-field approach, e.g.\ \cite{Abbott1981189}. To this
 end we introduce a background field $\bar A_\mu$ by splitting the
 full gauge field linearly into
\begin{equation}
  \label{eq:backsplit} 
  A_\mu=\bar A_\mu+a_\mu\,, 
\end{equation}
with $\langle a_\mu\rangle =0$. The background field enters the
background gauge-fixing condition
\begin{equation}
   \label{eq:backgauge}
   \bar D_\mu\, a_\mu =0\,,\quad {\rm with}\quad \bar D_\mu 
  =\partial_\mu + i g \bar A_\mu  \,.
\end{equation}
The background-field effective action $\Gamma_k[\bar A,\phi]$ depends
on the background field $\bar A_\mu$ and the fluctuation field
$\phi=(a_\mu,\, c,\,\bar c,\, ...)$. This setting allows for the
definition of a gauge invariant effective action
\begin{equation}
  \label{eq:Gamgauge} 
  \Gamma_k[A_\mu]=\Gamma_k[A_\mu,0]\,,  
\end{equation}
that is invariant under the gauge transformation 
\begin{equation}
  \label{eq:gaugetrafo} 
  A_\mu \to -\0{i}{g} ( U D_\mu U^{-1})\,,\quad {\rm with} \quad D_\mu
  = \partial_\mu + i g A_\mu\,.
\end{equation}
For constant fields this reduces to $A_\mu\to U A_\mu U^{-1}$ as for
$\varphi$ in \eq{eq:varphiP}. The related effective background
potential is given by
\begin{equation}
  \label{eq:Vglue} 
  V[A_0]= \01{\beta {\cal V}}\, \Gamma[A_0]\,,
\end{equation}
where $\cal{V}$ the spatial volume. It is evident from the discussion
in the last section that \Eq{eq:Vglue} simply is the potential for
constant $\varphi$ with $V(\varphi) = V[A_0(\varphi)]$ with the
relation \eq{eq:varphi} for $SU(3)$.

In perturbation theory the effective potential $V[A_0]$ \eq{eq:Vglue}
has first been computed in \cite{Gross:1980br,Weiss:1980rj}. It
features only the deconfining minimum at $A_0=0$.  A non-perturbative
approach for the computation of $ V[A_0]$ has been established in
\cite{Braun:2007bx, Marhauser:2008fz, Fister:2013bh}. By now
computations of the glue potential from a variety of methods have been
put forward, including the FRG and Dyson-Schwinger equations,
2PI-schemes, perturbative approaches and the lattice
\cite{Braun:2007bx, Braun:2009gm, Braun:2010cy, Marhauser:2008fz,
Heffner:2015zna, Reinhardt:2012qe, Fister:2013bh,
Fischer:2013eca, Reinhardt:2013iia, Fukushima:2012qa,
Reinosa:2014ooa, Reinosa:2014zta, Diakonov:2013lja,
Greensite:2012dy, Langfeld:2013xbf}.

The expectation value $\bar\varphi= g\beta \langle A_0\rangle/(2 \pi)$
is then given by the minimum of the background effective potential,
i.e. it is defined via
\begin{equation}
  \label{eq:<A>} 
  \left.  \0{\partial V[A_0]}{\partial A_0}\right|_{A_0=\langle
    A_0\rangle}=0\,,\quad {\rm with}\quad L[\langle A_0\rangle]\geq 0\,.
\end{equation}
As can be seen from \fig{fig:VglueContours}, the minimum is not
unique, and we choose the one which gives a real, positive
semi-definite traced Polyakov loop via \Eq{eq:LA0}, cf. our discussion
in \sec{sec:OrderParams}. The resulting critical temperature for $SU(3)$ is in
quantitative agreement with the lattice result with a critical
temperature $T_c/\sqrt{\sigma}=0.655$ in units of the string tension
$\sigma$, see \cite{Fister:2013bh}. Here, the biggest systematic
uncertainty concerns the relative scale setting in the continuum and
on the lattice, leading to a systematic error estimate of about
$10\%$ \cite{Fister:2013bh}.

Note that the two order parameters derived from the Polyakov loop, $\lLA$ and
$L[\lA0]$, are indeed closely related: Jensen's inequality yields the
relation \cite{Braun:2007bx} 
\begin{equation}
  \langle L[A_0]\rangle \leq L[\lA0]\,,
  \label{eq:Jensen}
\end{equation}
in the region where $L[A_0]$ is convex. Furthermore we have,
cf. \cite{Marhauser:2008fz, Fister:2013bh},
\begin{equation}
  L[\lA0]=0\Leftrightarrow \lLA=0\,.
  \label{eq:PLTc}
\end{equation}

Both these observables have been studied extensively and naturally
reproduce the order of the phase transition as well as the critical
temperature. However, the computations also reveal that the inequality
\eq{eq:Jensen} is far from being saturated. This statement holds for
$T> T_c$, i.\,e. in the center broken phase in pure Yang-Mills theory,
and for all temperatures in QCD.

For QCD effective models, e.\,g. \cite{Fukushima:2003fw,
  Megias:2004hj, Ratti:2005jh, Mukherjee:2006hq, Roessner:2006xn,
  Schaefer:2007pw, Fukushima:2008wg, Herbst:2010rf, Skokov:2010uh,
  Skokov:2010wb, Schaefer:2011ex, Herbst:2013ufa, Herbst:2013ail,
  Haas:2013qwp} this is particularly important.  There, one usually
relies on input from other methods to fix the gauge sector. The common
procedure is to introduce a Polyakov-loop potential $V_L(L,\bar L)$,
with $\bar L$ being the conjugate Polyakov loop. The coefficients of
$V_L$ are then fitted to lattice Yang-Mills data, i.e. the potential
is based on $\lL$\,.

This has to be contrasted with the fact that most low-energy effective
models can be derived from continuum QCD in a standard formulation
with gauge fields. Hence, they should be formulated in terms of
$\langle A_\mu\rangle$. Then, also the ultraviolet parameters of these
models at their initial scale, $\Lambda\approx 1$~GeV, can be
determined from continuum QCD at this scale, and the gauge sector is
related to the glue potential, $V[A_0]$.

Alternatively, a formulation in terms of $\lL$ is possible. Then,
however, the matter part of the models is subject to an inherent
Gau\ss ian approximation
\begin{equation}
  \label{eq:Gauss} 
  \langle L_1 \cdots L_n \bar L_1 \cdots \bar L_m\rangle = 
\prod_{i=1}^n \langle
  L_i\rangle \prod_{j=i}^m \langle \bar L_i\rangle\,, 
\end{equation} 
with the shorthand notation $L_i=L(\vec x_i)$ and similarly for $\bar
L_j$. Further reductions such as
\begin{equation}
  \label{eq:Gausstr} 
  \0{1}{N^2} \langle\Tr P_1 P_2 \rangle = \langle L_1 
  \rangle \langle L_2
  \rangle \,,
\end{equation}
with $P_i=P(\vec x_i)$ are used. \Eq{eq:Gausstr} neglects spatial
fluctuations of the Polyakov-loop variable $P(\vec x)$. Note that in
the Gau\ss ian approximation underlying \eq{eq:Gauss} and
\eq{eq:Gausstr}, the inequality \eq{eq:Jensen} saturates, which is
most easily seen in the Polyakov gauge \Eq{eq:polgauge}. In other
words, the difference between the observables $\lL$ and $L[\lA0]$
encodes the non-Gau\ss ianity of the Polyakov loop, and the explicit
results show the failure of the Gau\ss ian approximation. Thus, to
achieve a consistent treatment of the gauge sector in QCD effective
models, a reliable resolution of the deviation between these two
quantities is of utmost importance. In the present work we fill this
gap and compute both definitions of the Polyakov loop in a unified
approach from the FRG.

Apart from its relevance for effective models, the calculation of
$\lL$ from the FRG represents one more observable that is available
both, in the continuum and on the lattice, and allows us to
demonstrate that the relevant dynamics of the system is properly
accounted for in the functional continuum approach.

%%%%%%%%%%%%%%%%%%%%%%%%%%%%%%%%%%%%%%%%%%%%%%%%%%%%%%%%%%%%%%%%%%%%%%%%%%%%%%%%%%%%%%%%%%%%%%%%%%%%
%% Sec:PL Flows
%%%%%%%%%%%%%%%%%%%%%%%%%%%%%%%%%%%%%%%%%%%%%%%%%%%%%%%%%%%%%%%%%%%%%%%%%%%%%%%%%%%%%%%%%%%%%%%%%%%%
\section{Continuum approach to confinement order
  parameters}\label{sec:PLFlows}
As already emphasised, the Polyakov loop $\lLA$ is a non-local and
all-order correlation function of the temporal gauge
field. First-principle continuum approaches to QCD, however, are based
on the computation of local correlation functions of the
fundamental fields, and in particular $A_0$. It is this difference
that makes the continuum computation of $\lLA$ so difficult. At large
momentum scales $k\gg \Lambda_{\text{\tiny QCD}}$, this problem is
resolved within the perturbative expansion that allows us to drop the
higher-order correlation functions due to the small gauge coupling.
An extension of such a study into the non-perturbative domain is made
feasible by functional methods. In the present work we resort to the
functional renormalisation group (FRG).

\begin{figure}
  \includegraphics[height=.085\textheight]{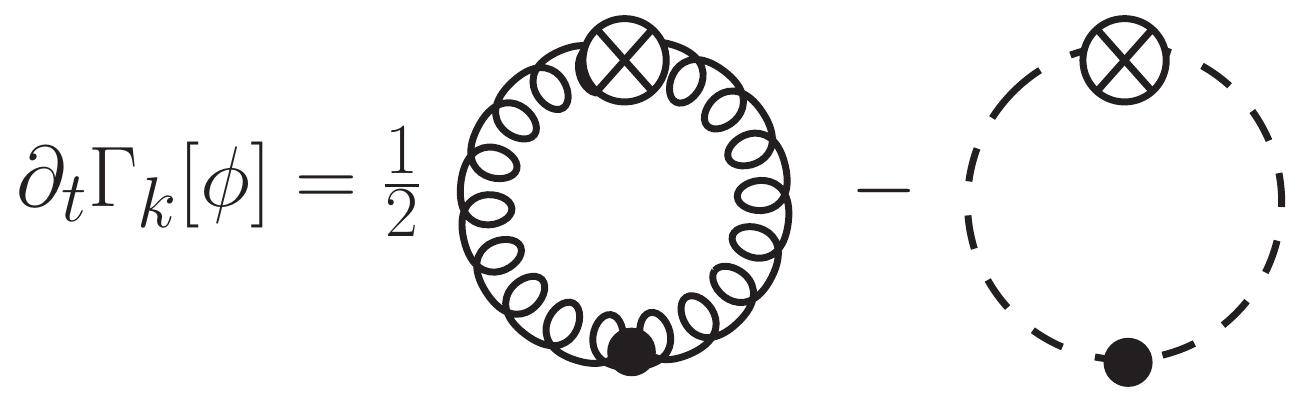}
  \caption{Diagrammatic representation of the FRG flow for YM
    theory. Curly lines denote gluonic degrees of freedom, while
    dashed lines represent the ghost and anti-ghost. The black dot
    indicates the full propagator and the crossed circle symbolises
    the regulator insertion.}
  \label{fig:YMFlow}
\end{figure}

\subsection{Functional renormalisation group}
The FRG allows us to systematically compute the scale dependence of
correlation functions. To this end, an infrared regulator function,
$R_k$, is introduced into the propagators of the theory. Such a
regulator suppresses the propagation of quantum and thermal
fluctuations below the infrared cutoff scale $k$. Lowering the cutoff
scale $k$ implements the Wilsonian idea of integrating out
fluctuations momentum shell by momentum shell. In the presence of the
scale $k$ the scale-dependent effective action, $\Gamma_k$, only
carries the quantum and thermal physics of momentum fluctuations above
the cutoff scale.  In turn, all quantum and thermal fluctuations below
this scale are suppressed. Hence $\Gamma_k$ interpolates between the
bare action at an ultraviolet scale $\Lambda$, and the full quantum
effective action for $k=0$. Its evolution is described by the
Wetterich equation \cite{Wetterich:1992yh},
\begin{equation}
  \partial_t\Gamma_k[\phi] = \012 {\Tr}\, G_k[\phi] \dot R_k\,,\ \ 
  {\rm with}\ \  G_k[\phi]=\0{1}{\Gamma^{(2)}_k[\phi] + R_k}\,,
  \label{eq:FRG}
\end{equation}
and the trace sums over momenta, internal indices and species of
fields, including minus signs for fermionic loops. In \eq{eq:FRG},
$\dot{R}_k = \partial_t R_k=k\partial_k R_k$ denotes the scale
derivative of the regulator w.r.t. the FRG time, $t=\log(k/\Lambda) $,
and $G_k$ is the full propagator at the scale $k$.  A pictorial
representation of this equation is given in \fig{fig:YMFlow} for the
special case of Yang-Mills (YM) theory used in this work. The curly
line denotes the scale-dependent gluon propagators, while the dashed
line represents the ghost propagator. The crossed circle symbolises
the regulator insertion, $\dot{R}_k$. By virtue of the regulator
function, \eq{eq:FRG} is ultraviolet and infrared finite. It has a
one-loop structure, but is fully non-perturbative.  The solution of
this equation relies on the propagators, which themselves obey flow
equations derived from \eq{eq:FRG}, see \cite{Fister:2011uw, Fister:2013bh}.

As can be seen from relation \eq{eq:Vglue}, the flow of the glue
potential, $V_k[A_0]$, is directly related to that of the effective
action,
\begin{equation}
  \partial_tV_k[A_0] = \frac{1}{\beta\mathcal{V}}\,\partial_t\Gamma_k[A_0]\,.
  \label{eq:FlowVk}
\end{equation}

Furthermore, in \cite{Pawlowski:2005xe} a flow equation for general
observables $I_k= \langle \hat I_k[J,\hat\phi] \rangle$ has been
derived, where $\hat\phi$ is the quantum field with $\phi=\langle
\hat\phi\rangle$. The flow equation for $I_k$ reads
\begin{align}
  \partial_t I_k[\phi] = -\frac12 {\rm Tr}\left\lbrace \left( G_k\dot
      R_kG_k\right)_{\bf ab}
    \frac{\delta}{\delta\phi_{\bf b}}\frac{\delta}{\delta\phi_{\bf a}}
    I_k[\phi]\right\rbrace\,.
  \label{eq:MasterFlowEq}
\end{align}
The bold letter indices ${\bf a},{\bf b}$ collect species of fields,
Lorentz and internal indices and again, $G_k$ denotes the full
propagator at RG-scale $k$. We refer the reader to
\cite{Pawlowski:2005xe} for more details and the derivation of
\eq{eq:MasterFlowEq}, and to \cite{Litim:1998nf, Berges:2000ew,
  Polonyi:2001se, Pawlowski:2005xe, Gies:2006wv, Schaefer:2006sr,
  Rosten:2010vm, Braun:2011pp, vonSmekal:2012vx} for QCD-related
reviews of the FRG.

\begin{figure}
  \includegraphics[height=.085\textheight]{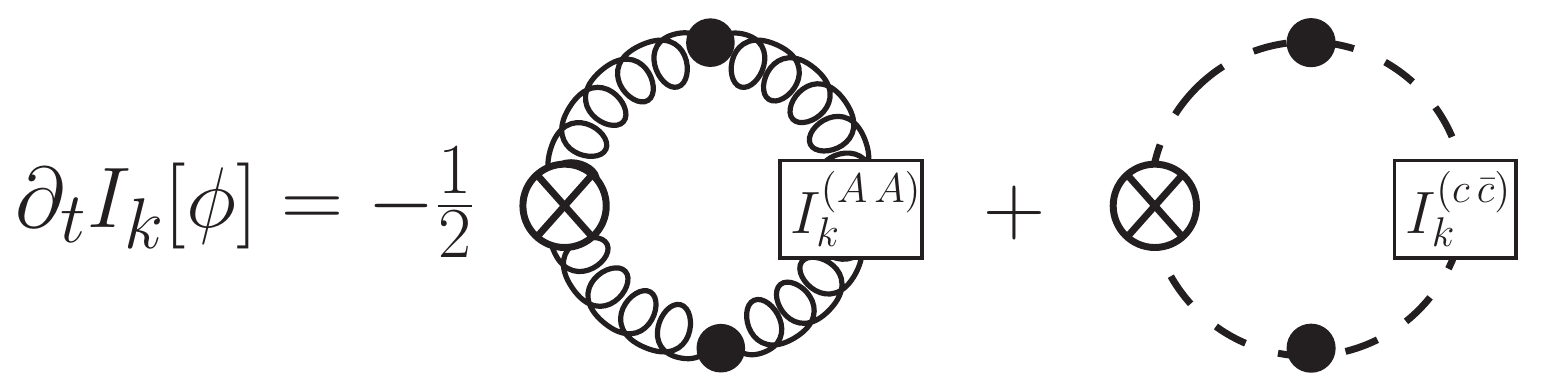}
  \caption{Diagrammatic representation of the FRG flow
    \eq{eq:MasterFlowEq} for general operators $I_k[\phi]$ in YM
    theory. The boxes indicate the second functional derivative
    of the operator w.r.t. the gluon and ghost fields, respectively.}
  \label{fig:IkFlow}
\end{figure}
\fig{fig:IkFlow} shows the diagrammatic representation of the flow
\eq{eq:MasterFlowEq}, again for YM theory. Its structure is similar to
\fig{fig:YMFlow}, except for the presence of second derivatives of the
operator $I_k$ w.r.t. the fields, indicated by the boxes. Once
more the propagators of the theory serve as input for the solution of
this flow equation.

The set of operators $I_k$ for which \eq{eq:MasterFlowEq} is valid,
includes, e.g., the (connected and disconnected) $n$-point correlation
functions $ \lbrace \langle
\phi_1\phi_2\dots\phi_n\rangle\,|\,n\in\mathbb N\rbrace$ as well as
$\tfrac{\delta\Gamma}{\delta\phi}$, cf. \cite{Pawlowski:2005xe}. In
particular, the observable $I_k~=~\lLA$ falls into this class of
operators.

\subsection{Flow equation for the Polyakov loop}\label{sec:flowPol}
From now on we restrict our discussion to $SU(3)$ Yang-Mills theory,
full QCD will be discussed elsewhere. We use the covariant
background-field formalism \cite{Abbott1981189} with the background
gauge-fixing condition \eq{eq:backgauge} in the Landau-deWitt
gauge. The vertices $\Gamma^{(n)}$ of the effective action
$\Gamma_k[A_\mu]$ are directly related to S-matrix elements, and hence
\eq{eq:Gamgauge} directly carries physics information. Its flow
equation is simply that of $\Gamma_k[A_\mu,0]$ in \eq{eq:FRG},
\begin{equation}
  \label{eq:FRGback}
  \partial_t\Gamma_k[A_0] = \012 {\Tr}\, G_k[A_0] \dot R_k\,,
\end{equation}
with 
\begin{equation}
  \label{eq:FRGbackG}
  G_k[A_0] = \0{1}{\0{\delta^2 \Gamma_k}{\delta\phi^2} +  
 R_k}[A_0,\phi=0]\,. 
\end{equation}
Eqs.~\eq{eq:FRGback} and \eq{eq:FRGbackG} entail that the propagators
of the fluctuation fields evaluated at vanishing fluctuation field,
$G_k[A_0]$, govern the flow of the gauge-invariant effective
action. This is the peculiar situation that the flow of a gauge
invariant quantity is governed by a gauge variant quantity, the
fluctuation field propagator. The latter satisfies a complicated
Slavnov-Taylor identity, see e.g.\ \cite{Pawlowski:2005xe}. However,
for vanishing fluctuation field, $\phi=0$, the propagator,
\eq{eq:FRGbackG}, is gauge covariant and can be expanded in gauge
covariant tensors. It is this property that enables us to use the
results of the Landau gauge, corresponding to the background
Landau-deWitt gauge \eq{eq:backgauge}, \cite{Braun:2007bx,
Braun:2010cy, Fister:2011uw, Fister:2013bh}. The gauge covariance
allows us to expand the two point correlator of the fluctuation field,
$\delta^2/\delta\phi^2\Gamma[A_0,0](p^2)$ in Landau-deWitt gauge about
the Landau gauge propagator \cite{Braun:2007bx,Fister:2013bh},
\begin{equation}
  \label{eq:expandLandau}
  \0{\delta^2\Gamma[A_0,0]}{\delta\phi^2}(p) =
  \0{\delta^2\Gamma_{\text{\tiny{Lan}}}}{\delta\phi^2}(D) + f(D,F,A_0)  \,,
\end{equation}
where the covariant tensor $f$ satisfies $f(D,0,0)\equiv 0$ and the
$A_0$-dependence of $f(D,F,A_0)$ indicates the dependence on the
covariant tensor $P(\vec x)$.  The subscript 'Lan' indicates the
standard Landau gauge with $\partial_\mu A_\mu=0$. The extension to
covariant momenta is done such that $ D\cdot \delta^2/\delta A_0^2
\Gamma_{\text{\tiny{Lan}}}(D) =0$, up to the gauge fixing term. For
constant fields the field-strength tensor vanishes, $F=0$, and the
covariant tensor reduces to $f(D,A_0)$.

\Eq{eq:expandLandau} allows us to use the results of the finite
temperature ghost and gluon propagators of \cite{Fister:2011uw} for
the first term on the right-hand side of \eq{eq:expandLandau}. The
computation of $f(D,F,A_0)$ will be discussed in
\sec{sec:input}. The Landau gauge two point functions at
finite temperature are conveniently parametrised as
\begin{equation}
  \label{eq:G2}
  \Gamma_{L/T,k}^{(2)}(p) = Z_{L/T,k}(p)\, p^2 P^{L/T} \,, \quad 
  \Gamma_{\bar c c,k}^{(2)}(p) = Z_{c,k}(p)\, p^2 \,,
\end{equation}
where $P^{L}$ and $P^T$ are the projection operators onto the
chromo-electric and chromo-magnetic gluon modes, and the identity
matrix in colour space is implied. For more details see
\cite{Fister:2011uw}. The following computation of Polyakov loop
expectation values only implicitly depends on $Z_T$, which relates to
the spatial components of the gauge field. For the sake of
simplicity we hence denote 
\begin{equation}
  \label{eq:Za}
  Z_{a,k}(p)  = Z_{L,k}(p)\,, \quad \quad Z_{\bar A,k}(p) = (Z_{\bar A})_{L,k}(p)\,. 
\end{equation}
Both, the fluctuating, $a_0$, and the background, $\bar A_0$, wave-function renormalisations are the 
chromo-electric ones.

With this choice of gauge, the background temporal gauge field,
$\bA0$, is in the Polyakov gauge. For vanishing fluctuation field,
$a_\mu=0$, we define, cf. Eqs.~\eq{eq:barvarphi} and \eq{eq:varphi},
\begin{equation}
    \label{eq:varphiold}
    \beta \bar g_k\, Z_{\bar A,k}^{1/2}\, \bar A_0 = 2\pi\left(\bar \varphi_3t^3 +
    \bar \varphi_8t^8\right) =: 2\pi \bar\varphi\,,
\end{equation}
where we have made the wave-function renormalisation $Z_{\bar
  A,k}=Z_{\bar A,k}(p_0=0,\vec p^2=k^2)$ of the background field
explicit. The wave-function renormalisation is evaluated at $\vec p\,^2
=k^2$, which is the important momentum scale in the flow. The running
coupling $\bar g_k$ is the background running coupling, evaluated at
the symmetric point $p^2 =k^2$. It tends toward a finite value in the
infrared and shows an infrared plateau, see e.g.\
\cite{Eichhorn:2010zc}.
\begin{figure*}
  \includegraphics[height=.26\textheight]{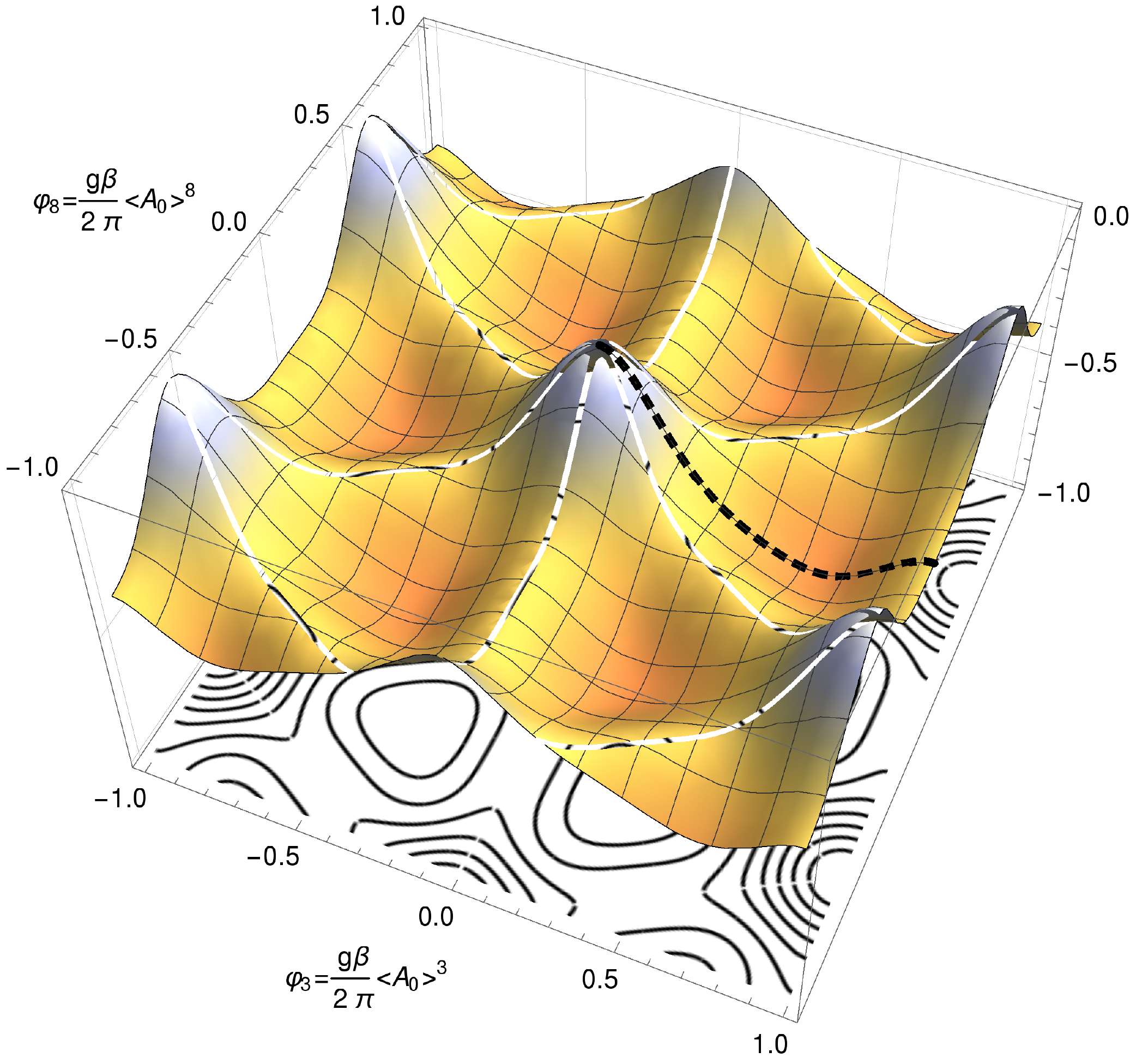}\hspace{20pt} 
  \includegraphics[height=.26\textheight]{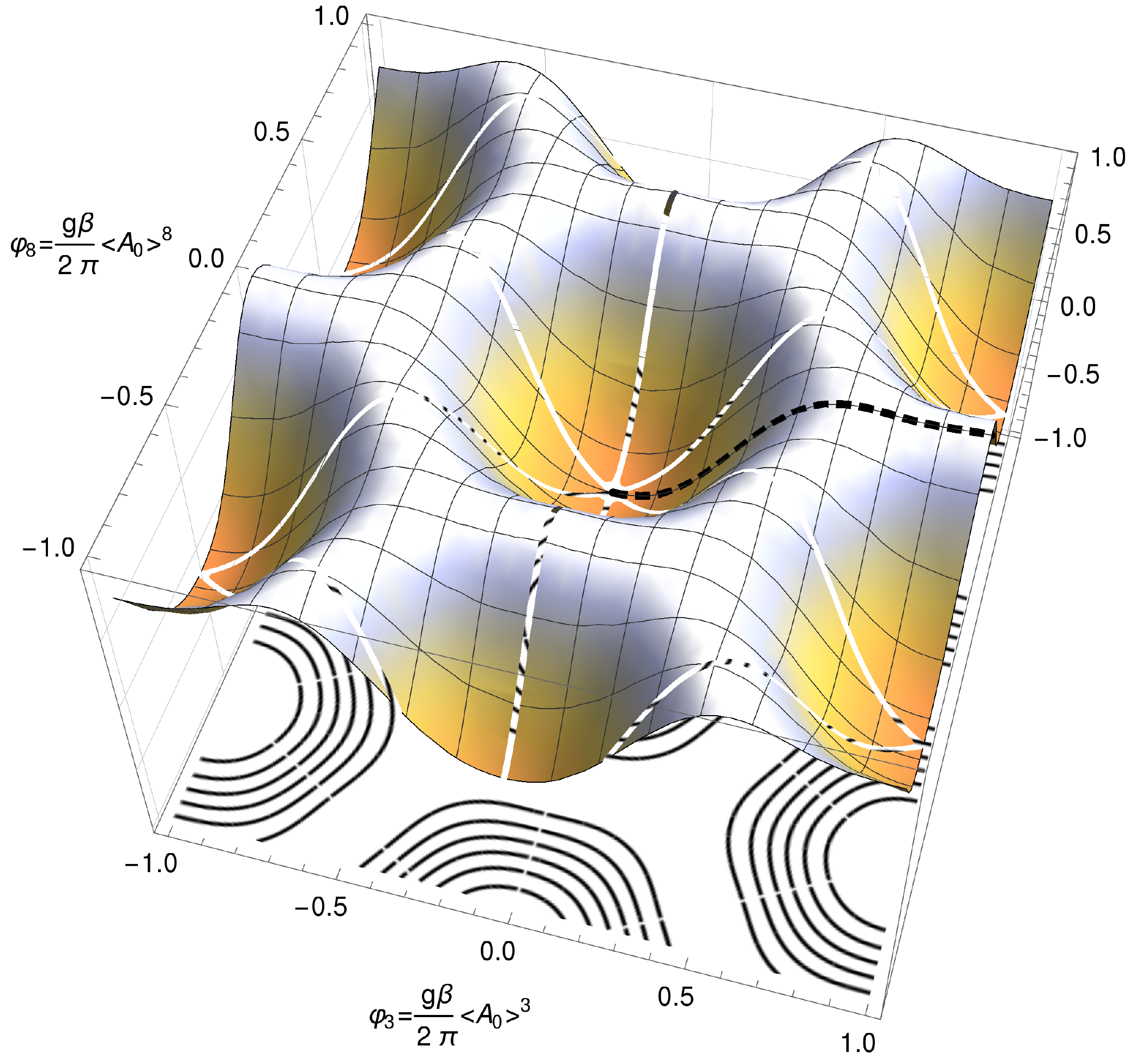}
  \caption{The infrared glue potential, $V(\varphi_3,\varphi_8)$, is
    shown in the confined phase (left, $T=236$~MeV) and in the
    deconfined phase (right, $T=384$~MeV). The periodicity of the
    potential, discussed in \sec{sec:OrderParams} is obvious. We
    restrict ourselves to the line $\varphi_8=0$ and $\varphi_3\geq0$
    (indicated by the black, dashed line), where one of the equivalent
    minima is always found, and where $L[\lA0]$ is
    real and positive semi-definite, cf. \fig{fig:VkSU3}.}
  \label{fig:Vglue3D}
\end{figure*}

Within our approach, the dependence on spatial gauge fields and ghosts
originates solely from $A_0-\vec A$ and $A_0$-ghost vertices. Such
vertices entail that all quantities involved are fully dressed, which
is accounted for by using the running coupling, $g=\bar g_k$, in
\eq{eq:varphiold}. Furthermore, the coupling $\bar g_k$ in
\eq{eq:varphiold} is the renormalisation group (RG) invariant
background running coupling. The combination $Z_{\bar A,k}^{1/2}\,
\bA0$ is RG-invariant, too, and hence $\bar\varphi$ is
RG-invariant. Moreover, we also have $\partial_t (g_k\, Z_{\bar
  A,k}^{1/2}\, \bA0)=0$, and hence $\partial_t \bar\varphi=0$.

Now we use that all background correlations can be expressed
in terms of those of the fluctuations with the help of Nielsen
identities. In \eq{eq:varphiold} this is particularly simple for
the cutoff-independent combination $\bar g_k Z_{\bar A,k}$. We
rewrite it as
\begin{equation}
  \label{eq:barAAmap}
  \bar g_k Z^{1/2}_{\bar A,k} = g_k Z^{1/2}_{A,k} \,, \quad Z_{A,k} = 
  \0{1}{Z_{a,k} Z_{c,k}^2}\,,\quad 
  g_k = \0{\bar g}{Z_{a,k}^{1/2}  Z_{c,k}}\,,
\end{equation}
where $g_k$ and $Z_{A,k}$ are defined via fluctuating wave-function
renormalisations, $Z_{a/c,k}=Z_{a/c,k}(p_0=0,\vec p\,^2=k^2)$. Moreover,
$\bar g$ is the background coupling at $k$ with $Z_{a,k}^{1/2}
Z_{c,k}=1$. We emphasise that the splitting \eq{eq:barAAmap} is not
unique, but rather a convenient definition of $Z_{A,k}$, enabled by
the cutoff independence of the combination $\bar g_k Z_{\bar A,k}$. 

For non-vanishing temporal fluctuation, $a_0\neq 0$, the
parametrisation \eq{eq:varphiold} generalises to
\begin{equation}
  \label{eq:varphia}
  2 \pi \varphi = \beta g_k\, Z_{A,k}^{1/2}\, \left(\bar A_0 
    + \tfrac{Z_{a,k}^{1/2}}{Z_{ A,k}^{1/2}}\, a_0\right) \,.
\end{equation}
Note that the combination \eq{eq:varphia} is RG-invariant, but not
$t$-independent for $a_0\neq 0$. Moreover, it is the choice of
the splitting \eq{eq:barAAmap} and the relation of $Z_{A,k}$ to the
fluctuation wave-function renormalisations that leads to the simple
ratio $Z_{a,k}^{1/2}/Z_{ A,k}^{1/2}$ in \eq{eq:varphia}: taking
derivatives w.r.t.\ the fluctuation fields naturally leads to powers
of the wave-function renormalisation and the fluctuation coupling
$g_{k} Z^{1/2}_a$, see the vertex parametrisation
\cite{Fischer2009b} It has been shown in \cite{Mitter:2014wpa} that
this approximation of the vertices is quantitatively reliable. The
ratio is then given by
\begin{equation}
  \label{eq:barZ}
  \tfrac{Z_{a,k}^{1/2}}{Z_{\bar A,k}^{1/2}} = \0{1}{Z_{c,k}}\,.
\end{equation}
With these prerequisites we can now apply the flow
\eq{eq:MasterFlowEq} to $\lLA$.  To this end we employ the
parametrisation
\begin{equation}
  \lLA = Z_{L,k}[\bar A,\phi]\cdot L(\varphi)\,, 
  \label{eq:lLpara}
\end{equation}
where the relative factor, $Z_{L,k}[\bar A,\phi]$, depends on all
fluctuation fields $\phi$ and on the background field $\bar
A$. Furthermore we have $\phi = \phi(\varphi)$ via \eq{eq:varphia}. It
is the renormalisation factor of the composite operator $L$, in
analogy to the wave-function renormalisation $Z_\phi$ of the field
operators $\phi$.

The $t$-derivative on the left-hand side of \eq{eq:MasterFlowEq} is
taken at fixed background $\bar A_0$. As we have seen above, this
corresponds to $\partial_t \bar\varphi=0$, which implies $\partial_t
L(\bar\varphi)=0$ and leads to the flow equation
\begin{align}
  \partial_t \lL  = &  L \,\partial_t Z_{L,k}  \nonumber\\
   = & -\frac{g^2_k\, Z_{a,k}\,\beta}{8 \pi^2 } \int \0{d^3p}{(2 \pi)^3} 
  \left[ G_k\dot R_kG_k
  \right]_{ab} (\omega_0=0,\vec p)\, \nonumber\\[2ex]
   & \times \left( Z_{L,k}\, L^{ba} +  Z_{L,k}^b \,L^a +Z_{L,k}^a\,L^b +  
    Z_{L,k}^{ba}\, L\right)\,,
    \label{eq:flowL}
\end{align}
with gauge group indices $a,b=1,...,N^2-1$ in $SU(N)$. For the sake of
brevity we have dropped the $\varphi$-arguments on the right-hand side
and have introduced the notation
\begin{equation}
  \label{eq:diffnot} 
  X^a = \0{\partial X}{\partial {\varphi_a}} \,,\qquad X^{ab} =
  \0{ \partial^2 X}{\partial {\varphi_a}\partial\varphi_b} \,,
\end{equation}
with $X=L, Z_{L}$\,. The $x_0$-independence of $\lA0$, due to the
Polyakov gauge for the background field, entails that only the lowest
Matsubara frequency, $\omega_0=0$, is present in this
equation. Furthermore, in this gauge only the derivatives with
$a,b=3,8$ are non-vanishing in \eq{eq:flowL}. Upon taking the trace,
only the components of the propagators that carry no explicit
dependence on the background field remain. Hence, the only field
dependence on the right-hand side is that of $L(\varphi)$ and
$Z_{L,k}$. As the field-dependence of the latter is only
generated by the flow, we conclude that the whole field dependence
originates from $L(\varphi)$.  Note also that the above analysis also
holds for general constant backgrounds without gauge fixing. In
summary, the symmetry properties of $L(\varphi)$ under shifts and
center transformations are carried over to $\lLA$ via the flow
\eq{eq:flowL}.

\Eq{eq:flowL} effectively is a flow equation for the ratio $Z_{L,k}$
of the two order parameters, $L(\varphi)=L[\lA0(\varphi)]$ and
$\lL(\varphi)$ for general background $\varphi$. Eventually we
evaluate all observables at the minimum $\bar\varphi$ defined by
\eq{eq:<A>}. $Z_{L,k}$ provides a measure for the different impact of
quantum and thermal fluctuation on the two definitions of the Polyakov
loop. It hence is a measure of the non-Gau\ss ianity of Polyakov loop
correlations as well as of gauge field correlations that is relevant
for low energy effective models. The seeming parametric singularity of
\eq{eq:flowL} when $L(\varphi)=0$ is lifted by our parametrisation
\eq{eq:lLpara}, which guarantees that in this case also $\lL=0$\,, cf.
\Eq{eq:PLTc}.

It is also worth noting that the off-diagonal entries in the last line
of \Eq{eq:flowL} in general are non-vanishing and purely
imaginary. They couple to the off-diagonal terms of the propagator,
$G_k^{38},\, G_k^{83}$, which are only present at finite quark
chemical potential. This property facilitates the emergence of
different, real values for the Polyakov loop and its conjugate, $\lL$
and $\lLc$, at finite density. This is expected since these
observables relate to the free energies of quarks and anti-quarks,
respectively.

Additionally, knowledge of $Z_{L,k}$ enables us to translate between
the glue effective potential, $V(\varphi)$\,, cf. \Eq{eq:Vglue}, as
computed in \cite{Braun:2007bx, Braun:2009gm, Braun:2010cy,
Marhauser:2008fz, Heffner:2015zna, Reinhardt:2012qe, Fister:2013bh,
Fischer:2013eca, Reinhardt:2013iia, Fukushima:2012qa,
Reinosa:2014ooa, Reinosa:2014zta}, and the Polyakov loop potential
used in models, $V_{L}(L=\lL)$ by the above relation $\lL(\varphi) =
Z_L(\varphi) L[\varphi]$\,. This can, for example, be used to evaluate
and improve model potentials.

\subsection{Input}\label{sec:input}
In order to solve the flow \eq{eq:flowL}, the scale-dependent YM
propagators and the glue potential, $V_k(\varphi)$, are the only
inputs needed. The covariant tensor $f(D,A_0)$ is directly related to
the background potential via \eq{eq:varphia}, to wit
\begin{equation}
  \label{eq:fV}
  f(D,A_0)= Z_{a,k}\, \beta^2\0{g_k^2 }{4 \pi^2} V_k^{ab}(\varphi)\,. 
\end{equation} 
Here the prefactor $Z_{a,k}$ carries the correct RG-scaling of a
gluonic two-point correlation function of the fluctuation field, and
the rest is RG-invariant. FRG calculations for the YM propagators have
previously been put forward in \cite{Fister:2011uw}, and an FRG
approach to the glue potential has been presented in, e.g.,
\cite{Fister:2013bh}. Here we build upon these results to achieve a
consistent description of fluctuation effects on the Polyakov loop.

\begin{figure}
  \centering
  \includegraphics[width=\columnwidth]{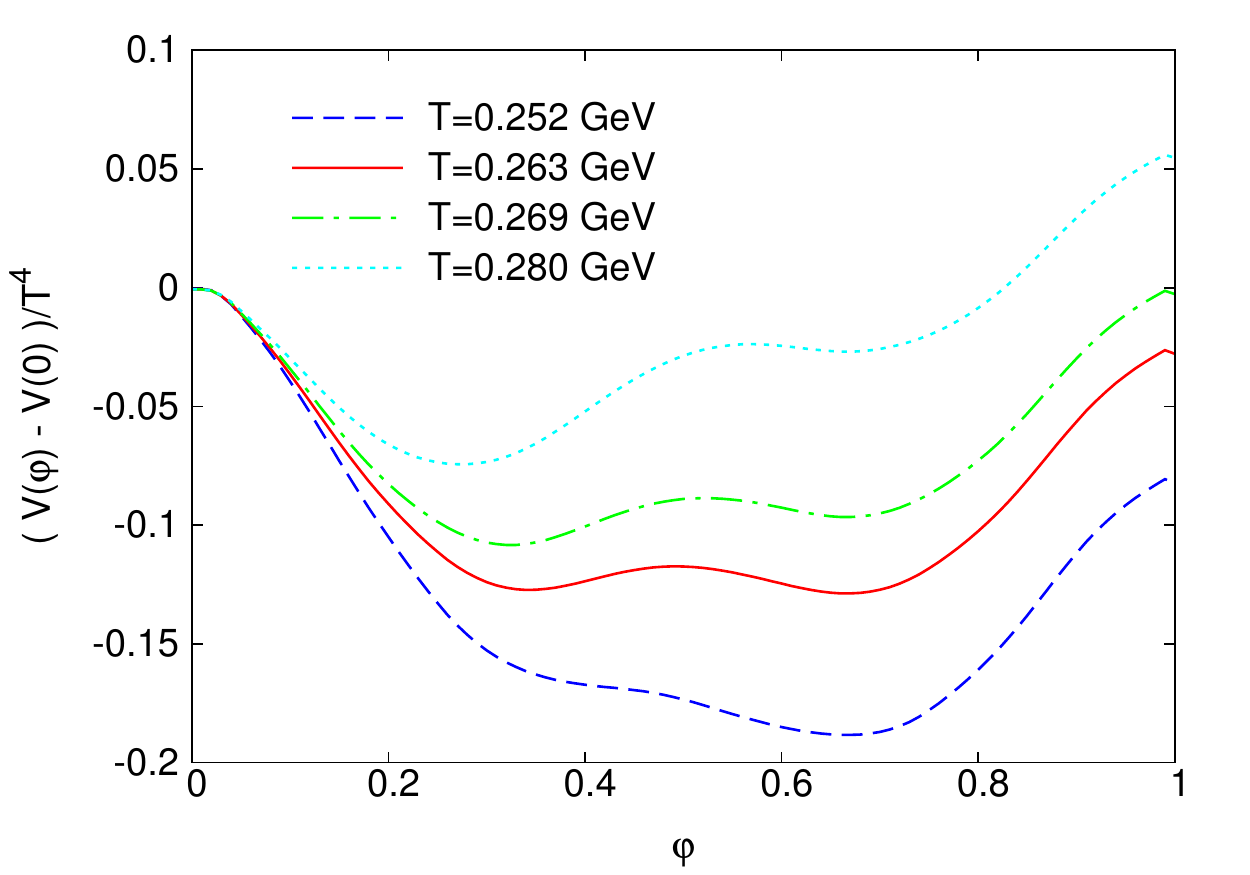}
  \caption{FRG result for the $SU(3)$ glue potential along the line
    $\varphi_3\geq0, \varphi_8=0$ for various temperatures. The
    critical temperature, associated with the first-order transition
    is given by $T_c=264\pm26$~MeV.}
  \label{fig:VkSU3}
\end{figure}

First, we briefly recapitulate the calculation of the glue potential,
whose flow equation is depicted diagrammatically in \fig{fig:YMFlow}.
To solve this equation we make use of the temperature- and
scale-dependent results for the ghost and gluon propagators from
\cite{Fister:2011uw}. There, results have been obtained
  with the ghost and gluon regulators
\begin{equation}
  \label{eq:regs}
  R_{a/c,k}(p^2) = \bar Z_{a/c,k}\, p^2\, r(p^2/k^2)\,, \quad \quad r(x)=
  \frac{x}{ e^{x^2}-1}\,,  
\end{equation}
which ensures the necessary momentum locality of the flow, as well as
an exponential thermal decay proportional to $k/T$ for large cutoff
scales. The wave-function renormalisations $\bar Z_{a/c}$ are related
to $Z_{a/c}$, for details see \cite{Fister:2011uw}. It has been shown
in \cite{Fister:2013bh} that the corresponding glue potential
correctly reproduces the order and critical temperature of the $SU(3)$
deconfinement phase transition. The resulting glue potential,
$V_{k\to0}(\varphi_3,\varphi_8)$, is shown in \fig{fig:Vglue3D} in the
confined (left) and deconfined phase (right). The symmetries discussed
in \sec{sec:OrderParams} are clearly visible and we have indicated the
line $\varphi_8=0, \varphi_3\geq0$, to which we restrict, by the
black, dashed line. The corresponding potential along this line is
then shown in \fig{fig:VkSU3} for several temperatures below and above
the critical temperature, $T_c=264\pm26$~MeV. As also discussed in
\cite{Fister:2013bh}, the absolute scale in this computation is set by
a comparison of the peak position in the propagators with the lattice
\cite{Maas:2011se, Fischer:2010fx, Maas:2011ez}, which results in an
error in $T_c$ of about $10~\%$.

The physical gauge field, $\lA0\sim\bar\varphi$, is determined as the
global minimum of the infrared glue potential, $V_{k\to0}$, see
\eq{eq:<A>}. The order parameter $L(\bar\varphi)$ is calculated via
\Eq{eq:LA0}. FRG data for the YM propagators are available up to
$T\lesssim1.1$~GeV \cite{Fister:2013bh}, which allows us to compute
the glue potential fully non-perturbatively up to this scale. For all
temperatures the glue potential $V(\varphi)$ can be fitted well via
the perturbative Weiss potential as
\begin{equation}
  V(\varphi) = a(T)\, V_W^{\rm SU(3)}(\varphi) 
  + b(T)\, V_W^{\rm SU(3)}(\varphi)^2\,,
\end{equation}
where the Weiss potential for $SU(2)$ is given by $V_W^{\rm
  SU(2)}(\varphi)=(1-(1-2\varphi)^2)^2$ and its $SU(3)$-counterpart
can be constructed in the standard way as
\begin{equation}
  V_W^{\rm SU(N)}(\varphi) = \012\sum_{n=1}^{N^2-1} 
  V_W^{\rm SU(2)}(x_n(\varphi))\,,
\end{equation}
with $x_n=\lbrace 0,
\pm\varphi_3,\pm\tfrac{\varphi_3\pm\sqrt{3}\varphi_8}{2} \rbrace$ in
$SU(3)$ see e.g. \cite{Braun:2010cy}. For high temperatures, $T\gtrsim
1$~GeV we find that the coefficient $b(T)$ is rather small and the
potential is well approximated by the perturbative form. We exploit
this observation in the following and replace the glue potential in
\eq{eq:fV} by the Weiss potential at $T\gtrsim1$~GeV.

Furthermore, it turns out that the thermal propagators can be
approximated by the $T=0$ propagators plus a Debye mass which is
fitted to the data. At large temperatures, $T\gtrsim1$~GeV, where no
FRG data for the propagators is available yet, we use this
construction.  We have checked that the detailed form of the mass term
has little impact on the resulting Polyakov loop. In the plots shown
in the next section, we indicate the scale where we switch from
first-principles data to the fit for the Debye mass and Weiss
potential by a dashed, vertical line.

To finally solve the flow \eq{eq:flowL} for $\lLA$, the last remaining
ingredient is the running coupling, $g_k$. Within our approach, it is
related to the strength of the ghost-gluon vertex, $g_k^2 \sim
Z_{a,k}^{-1} Z_{c,k}^{-2}$, and can also be deduced from the
propagators, cf.~\cite{Fister:2011uw}.

%

%%%%%%%%%%%%%%%%%%%%%%%%%%%%%%%%%%%%%%%%%%%%%%%%%%%%%%%%%%%%%%%%%%%%%%%%%%%%%%%%%%%%%%%%%%%%%%%%%%%%
% Sec: Renormalization 
%%%%%%%%%%%%%%%%%%%%%%%%%%%%%%%%%%%%%%%%%%%%%%%%%%%%%%%%%%%%%%%%%%%%%%%%%%%%%%%%%%%%%%%%%%%%%%%%%%%%
\section{Fluctuations and renormalisation}\label{sec:Renormalization}

Following the procedure outlined in the last section we are now in the
position to compute both Polyakov loops, $L(\bar\varphi)$ and $\lL =
Z_{L,k=0}(\bar\varphi) L(\bar\varphi)$. For the computation of the
latter we first remark that the composite operator $\lL$ is
UV-relevant and has to be renormalised.  In the present functional
renormalisation group approach the renormalisation procedure is easily
accessible. It simply entails that
\begin{equation}
  \label{eq:RenPL}
  \Lambda \partial_\Lambda \lL_{k=0} \stackrel{!}{=} 0\,, 
\end{equation}
for the renormalised traced Polyakov loop: the expectation value
$\lLA$ at vanishing cutoff scale, $k=0$, does not depend on the
initial cutoff scale, $k=\Lambda$. Hence, this amounts to adjusting
the $\Lambda$-dependence of $\lLA_\Lambda$. For \eq{eq:RenPL} to hold,
$\lLA_\Lambda$ has to satisfy the flow equation \eq{eq:flowL}, see
e.g.\ \cite{Pawlowski:2005xe}. Within the parametrisation
\eq{eq:lLpara} this translates into the requirement that
$Z_{L,\Lambda}$ has to satisfy \eq{eq:flowL}. In turn, the
unrenormalised traced Polyakov loop is defined by using the trivial
initial condition $Z_{L,\Lambda}=1$, which violates
\eq{eq:RenPL}. Thus, the multiplicative factor $Z_{L,\Lambda}$ carries
the renormalisation of the Polyakov loop $\lLA$, and is related to the
multiplicative renormalisation factor $Z_{\text{\tiny{lat}}}$ used on
the lattice.

For asymptotically large scales, that is $\Lambda/\Lambda_{\text{\tiny
    QCD}}\to \infty$ and $\Lambda/T_{\text{\tiny{max}}}\to \infty$, it
is sufficient to solve \eq{eq:RenPL} up to sub-leading terms that
vanish in this limit. Asymptotically large $
\Lambda/\Lambda_{\text{\tiny QCD}}$ guarantees the perturbative limit
that facilitates the computation of the initial $Z_{L}$. The second
requirement, $ \Lambda/T_{\text{\tiny{max}}}\to\infty$ guarantees the
temperature-independence of the initial condition.

In the above asymptotic limit, the temperature dependence of the gluon
propagators, including the term proportional to the potential
curvature $V^{ab}$, and of the gluonic wave-function renormalisation
$Z_a$ decays exponentially with $\exp \{-c(r) k/T\}$, see
\cite{Fister:2011uw, Fister:2013bh,Fister:2015eca}. The prefactor
$c(r)$ depends on the shape function $r$ in \eq{eq:regs}, and vanishes
for non-analytic regulators, see \cite{Fister:2015eca}. In such a case
the thermal suppression is only polynomial in $T/k$.  For the present
regulators, \eq{eq:regs}, we have an exponential decay.  The same
exponential decay holds for general correlation functions, and hence
also applies to the coupling $\alpha_s$. Accordingly, the temperature
dependence of the flow \eq{eq:flowL} decays exponentially, reflecting
the temperature independence of the UV renormalisation. Moreover, the
flow also becomes $\varphi$-independent exponentially fast as the sum
of the $L^{ab}$-terms is proportional to $L$ for vacuum propagators
which are proportional to $\delta^{ab}$. Hence, all $Z_L^{a},Z^{ab}_L$
terms on the right-hand side can be dropped, and we are left with the
asymptotic flow
\begin{equation}
  \label{eq:flowLas}
  \0{\partial_t Z_{L,k}}{Z_{L,k}} = \frac{g^2_k\, Z_{a,k}\,\beta}{6
      } \int \0{d^3p}{(2 \pi)^3} \left[ G_k\dot R_kG_k
    \right]_{bb} (0,\vec p)\,,
\end{equation}
for the renormalisation factor. The prefactor has a logarithmic
scaling with $k$ from both, the running coupling $g_k^2$ and the gluon
wave-function renormalisation $Z_{a,k}$. The scaling of the latter is
cancelled by the logarithmic scaling of the propagators and
regulator. The remaining spatial momentum integral scales linearly
with the cutoff scale. Hence we arrive at the total scaling
\begin{equation}
  \label{eq:Zren}
  \frac{\partial_t Z_{L,k}}{Z_{L,k}} =   z_L\,\alpha_{s,k}\,\0{k}{T}\,,
\end{equation}
with the strong running coupling $\alpha_{s,k}=g^2_k/(4\pi)$ at $T=0$
up to exponentially suppressed terms. The factor $z_L$ is given by
\begin{equation}
  \label{eq:a}
  z_L = \frac{2 \pi}{ 3} \int \0{d^3p}{(2 \pi)^3} \0{1}{k}\left[
    Z_{a,k}\, G_k\dot R_kG_k \right]_{bb} (0,\vec p)\,,
\end{equation}
with the propagators $G_k$ and $Z_{a,k}$ at vanishing temperature, up
to exponentially suppressed terms. Moreover, $z_L$ tends towards a
constant for large cutoff scales, $z_{L,k\to\infty}= 0.0177$.

Note also that the exponential thermal decay does not hold in momentum
space at vanishing cutoff scale. All correlation functions, including
the propagator and the running coupling, $\alpha_s(p,T)$, only show a
polynomial decay of the thermal contributions with large momenta, that
is with powers of $T/p$, similarly to the polynomial decay with $T/k$
for non-analytic cutoffs. However, contrary to the latter, the
polynomial decay with $T/p$ relates to physics, for example that of
the propagator relates to the Tan relations in many-body physics, for
a discussion in Yang-Mills theory see \cite{Fister:2011uw}. In turn
this entails that the choice of temperature-independent
renormalisation conditions in a non-perturbative approach based on the
running coupling at asymptotic momentum scales is intricate. Hence, in
particular the identification of the finite temperature running
coupling at large momenta with the $T=0$ one, $\alpha_s(p\gg
T,T)=\alpha_s(p\gg T, 0)$ introduces a temperature-dependent
renormalisation scheme due to the sub-leading powers of $T/p$.

We emphasise that for the present renormalisation scheme with the
regulator \eq{eq:regs} this is avoided: at large cutoff scales $k/T
\gg 1$ the difference of the couplings $\alpha_{s,k}(p,T)$ and
$\alpha_{s,k}(p,T=0)$ is exponentially suppressed with
$f(T/k,T/p)\exp\{ - c_T k/T\}$. Here, $f(x,y)$ has power law
suppressions for both, $k/T\to\infty$ and $p/T\to\infty$. At $k=0$ the
exponential suppression disappears. Hence the integrated flow leads to
the sub-leading momentum-dependent terms in the coupling in $T/p$ that
are only power law suppressed at large momentum scales. This has been
shown in \cite{Fister:2011uw}.

\begin{figure}
  \centering
  \includegraphics[width=\columnwidth]{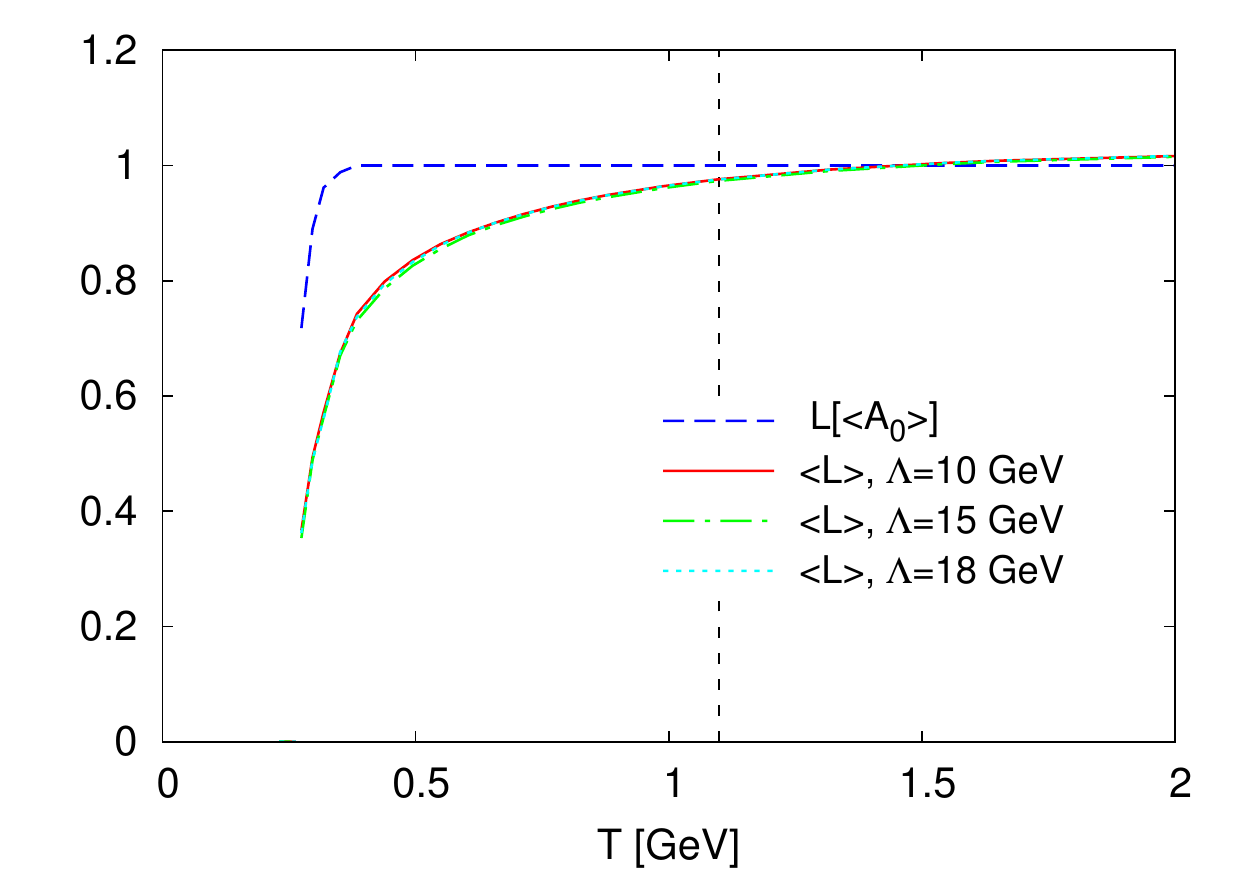}
  \caption{Demonstration of the cutoff-independence of the Polyakov
    loop, $\lL$ with $a_1=0$ in \eq{eq:ZLlambda}, and the comparison
    to $L[\lA0]$. The high-T limit in both cases is unity.}
  \label{fig:PL_ren}
\end{figure}

Nonetheless, a temperature-dependent renormalisation scheme is a
consistent choice, and such a general setting is taken into account in
the present setting by allowing for thermal shifts in the cutoff
scale. This amounts to a sub-leading change of the initial condition
for the flow of $Z_L$ proportional to $T/\Lambda$, and hence a
sub-leading contribution to the flow proportional to $T/\Lambda$. The
flow is peaked at about the temperatures scale, and we are led to the
final expression for the renormalisation factor $Z_{L,\Lambda}$ at
asymptotically large cutoff scales $\Lambda$
\begin{equation}
  \label{eq:ZLlambda}
  Z_{L,\Lambda}(T) = e^{a_{\text{\tiny{cont}}} \Lambda/T}\,, \quad
  a_{\text{\tiny{cont}}} =a_0 + a_1 \frac{T}{\Lambda}\,\alpha_s( c_T
  T) + \dots \,,
\end{equation} 
where the leading term $a_0 =1/\Lambda \int^\Lambda d k\,
z_L\,\alpha_{s,k}$ is proportional to $\alpha_{s,\Lambda}$ for large
cutoff scales. However, it is very slowly varying due to its only
logarithmic decay, and for the chosen initial cutoffs $\Lambda=10, 15,
18$~GeV in \fig{fig:PL_ren} it is given by $a_0=0.0175$. More
importantly, \fig{fig:PL_ren} demonstrates the renormalisation group
invariance of the traced Polyakov loop observable $\lL$. The
higher-order term, with free coefficient $a_1$, acts as a constant
shift in the Polyakov loop at low temperatures, and decreases at large
$T$, respecting the correct high-temperature limit. It implements the
subleading temperature-dependent renormalisation discussed above, and
the parameters $a_1, c_T$ specify the scheme. Note also that for small
temperatures above, but close to the phase transition the flow for
$Z_L$ is peaked at the mass gap of the gluons and the scale $c_T T$ in
$\alpha_s( c_T T) $ has to freeze at about the mass gap. This entails
that the sub-leading term related to the potential
temperature-dependent renormalisation provides a constant off-set in
the infrared, and the ratio for expectation values of the traced
Polyakov loop is independent of the temperature-dependent part of the
renormalisation. In summary, the potential temperature-dependent RG
scheme introduces in $Z_L$ a temperature-dependent, but
cutoff-independent factor for asymptotically large temperatures, while
it is constant for low temperatures due to the plateau in the
background coupling. In \fig{fig:PL_ren} we have used $a_1=0$, since
it does not explicitly depend on $\Lambda$. This subtlety is discussed
further in the next section.

The above renormalisation procedure in the continuum is to be compared
to the lattice renormalisation. There, it is known that taking the
infinite volume limit at a fixed temperature leads to a vanishing
Polyakov loop and an appropriate renormalisation procedure is needed
\cite{Kaczmarek:2002mc, Zantow:2003uh, Gupta:2006qm, Gupta:2007ax}.
It was suggested already by Polyakov \cite{Polyakov1980171}, that the
linear divergences appearing in $\lL$ at any order in perturbation
theory can be combined into an exponential factor.  This leads to the
definition of the renormalised Polyakov loop
\begin{equation}
  \lL_{\text{\tiny{lat}}} = Z_{\text{\tiny{lat}}}(T) \lL_{\text{\tiny{bare}}}\,,\quad 
  \text{with}\quad Z_{\text{\tiny{lat}}}(T) = e^{a_{\rm\tiny{lat}}\Lambda/T}\,, 
  \label{eq:renPL}
\end{equation}
where \eq{eq:ZLlambda} suggests
$a_{\text{\tiny{cont}}}=a_{\text{\tiny{lat}}}$ up to differences in
the renormalisation condition. On the lattice, the constant
$a_{\text{\tiny{lat}}}$ can be fixed via the zero-temperature
heavy-quark potential \cite{Kaczmarek:2002mc, Zantow:2003uh,
  Gupta:2006qm}, for other renormalisation schemes see
e.g. \cite{Gupta:2007ax, Dumitru:2003hp, Gavai:2010qd,
  Mykkanen:2012ri}. Such a renormalisation procedure is unique only up
to a constant, i.\,e. an overall multiplicative renormalisation
remains, cf. \cite{Gupta:2006qm, Gupta:2007ax}. The different
normalisations used on the lattice and in the continuum may also
introduce a temperature-dependent part of the renormalisation as
discussed above. This has to be taken into account in a comparison of
the different schemes, which is deferred to the next section.

While both continuum observables, $L[\lA0]$ and $\lL$, serve as order
parameters for the confinement-deconfinement transition,
\fig{fig:PL_ren} reveals clear differences between the two above
$T_c$. It had been found previously, and is also confirmed by our
data, that the overall shape of $L[\lA0]$ is much steeper than that of
$\lL$. The difference between these two observables is quantified by
$Z_{L,k\to0}(T)$ and is due to the different impact of thermal and
quantum fluctuations on both observables.

First we note that $L[\lA0]$ saturates at its high-temperature limit,
$L[\lA0]\stackrel{T\to\infty}{\longrightarrow}1$, already at $T\gtrsim
0.4$~GeV. In turn, $\lL$ rises much more gradually. It eventually
overshoots unity at $T\approx 1.5$~GeV and approaches its high-T limit
of one from above, in agreement with the expectation from perturbation
theory \cite{Gava1981285}.

The non-Gau\ss ianity of the Polyakov loop in terms of correlations of
the gauge field is quantified by the difference between $L[\lA0]$ and
$\lL$ as discussed in \sec{sec:OrderParamsgauge}. It is clearly
visible in \fig{fig:PL_ren}, and originates in two qualitatively
different mechanisms: The first one simply encodes the renormalisation
of the gauge field correlation functions that are in line with
(resummed) thermal perturbation theory. These deviations are similar
to those observed in the pressure, which are known to be significant
even for temperatures $T\gg T_c$ as the approach to the
Stefan-Boltzmann limit is rather slow. Still, they are captured well
by HTL or similar resummation schemes such as (perturbative)
two-particle-irreducible computations. This is visible in $Z_{L,k}$
which is approximately linear in the cutoff scale up to the thermal
gap for temperatures $T\gtrsim (2-3)T_c$. In turn, for temperatures
$T\lesssim (2-3)T_c$ the factor $Z_{L,k}$ shows an additional
significant drop at about the non-perturbative mass gap of Yang-Mills
and freezes below this scale.

In summary, the non-perturbative renormalisation factor $Z_L(T)$
carries much of the information about the perturbative and
non-perturbative thermal and quantum fluctuations. The deviations are
particularly large for the low temperature regime $T\leq (2-3)
T_c$. In this regime we also see a significant non-trivial dynamics
related to the confinement-deconfinement phase transition. It is
expected that these differences have a significant impact on model
calculations if the standard Gau\ss ian approximation is lifted.

\section{Continuum and lattice results for
  $\lLA$ and renormalisation}\label{sec:Comparison}
\begin{figure}
  \centering
  \includegraphics[width=\columnwidth]{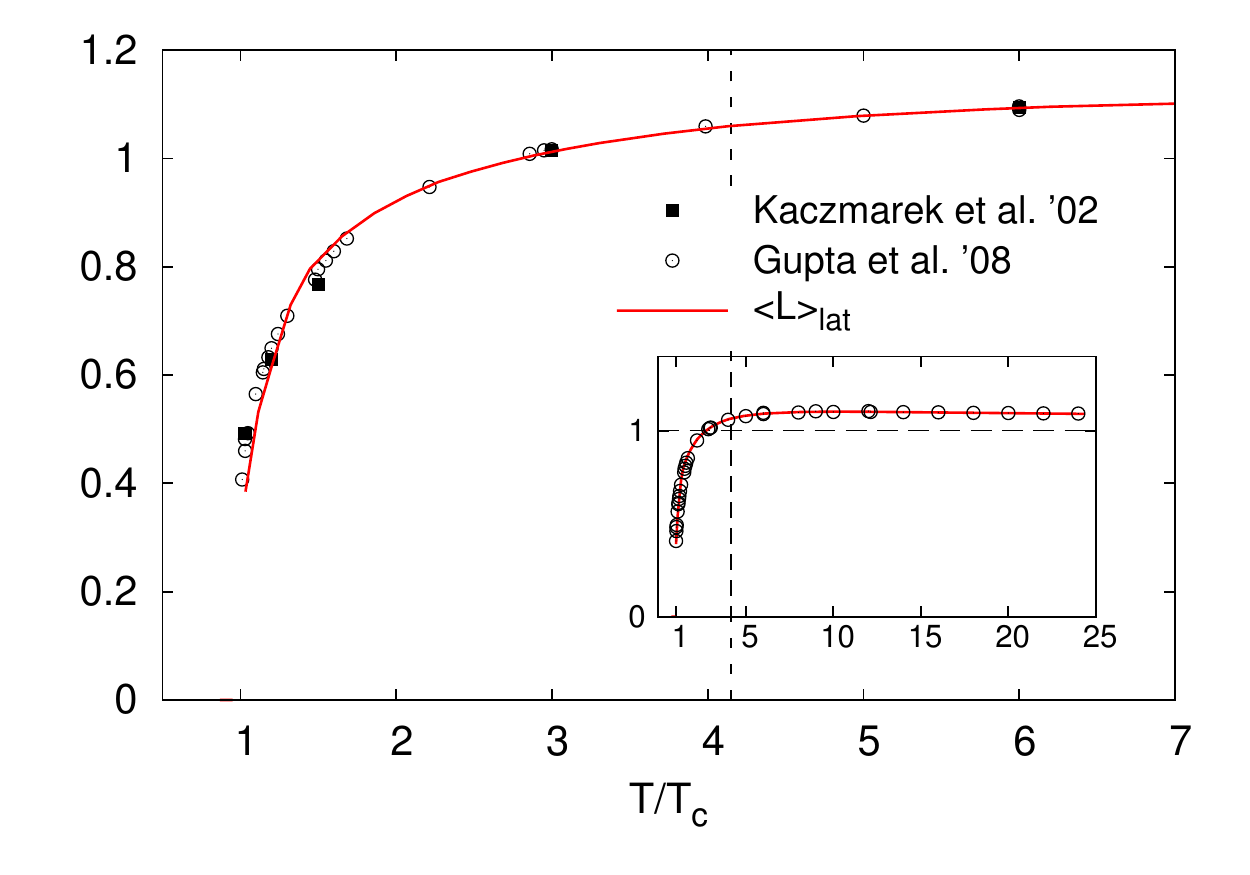}
  \caption{Polyakov loop, $\lL$, compared to the renormalised lattice
    result, \cite{Kaczmarek:2002mc, Gupta:2007ax}.  The inset shows
    the high-T behaviour vs $T/T_c$.}
  \label{fig:PLlat_T}
\end{figure}

Now we turn to a comparison of $\lL$ in the continuum and on the
lattice. This is done as a function of the reduced temperature,
$T/T_c$, to accommodate for the differences in the relative scale
setting procedure mentioned before. Using the renormalisation
procedure discussed above with only $a_0$, the ratio of the continuum
and lattice Polyakov loop expectation values is given by
$\lL_{\text{\tiny{cont}}}/\lL_{\text{\tiny{lat}}}= 0.92$ for
temperatures $T/T_c \lesssim 10$, that is up to about $3$~GeV. It has
been discussed at length in the last section that such a constant low
temperature off-set is introduced by a relative temperature-dependent
renormalisation. Hence, this result confirms the quantitative
agreement of both computations. Above this temperature we observe a
change of this ratio towards unity triggered by the normalisation of
both Polyakov loops at infinite temperature.  As discussed in
\sec{sec:Renormalization}, the subleading term in the continuum
renormalisation, that signals a temperature-dependent renormalisation
scheme, on the other hand, is approximately constant at low $T\lesssim
10 T_c$. With the choice $a_1=0.47$ it accounts for the factor $1.088
=1/0.92$, see \fig{fig:PLlat_T}. Note that this term also accommodates
possible sub-leading corrections of the continuum computation in the
presence of full gluon propagators for asymptotically large
temperatures. In the present work we have tested a large, but not
complete set of sub-leading temperature corrections to the
propagators, and did not observe any impact. In summary this strongly
suggests a relative temperature-dependent renormalisation. To
highlight the quantitative agreement up to asymptotically high
temperatures, the inset in \fig{fig:PLlat_T} shows the continuum
Polyakov loop with full renormalisation \eq{eq:renPL}, and the lattice
Polyakov loop from \cite{Gupta:2007ax} vs. $T/T_c$. A full
investigation of this large temperature intricacy for $T/T_c \gg 1$
will be discussed elsewhere.

%

%%%%%%%%%%%%%%%%%%%%%%%%%%%%%%%%%%%%%%%%%%%%%%%%%%%%%%%%%%%%%%%%%%%%%%%%%%%%%%%%%%%%%%%%%%%%%%%%%%%%
% Sec: Conclusion and Outlook
%%%%%%%%%%%%%%%%%%%%%%%%%%%%%%%%%%%%%%%%%%%%%%%%%%%%%%%%%%%%%%%%%%%%%%%%%%%%%%%%%%%%%%%%%%%%%%%%%%%%
\section{Conclusion and Outlook}\label{sec:conclusion}

In this work we have discussed different order parameters for the
confinement-deconfinement transition in Yang-Mills theory, based on
the Polyakov loop variable $P(\vec x)= \exp(2 \pi i \varphi)$. The
common order parameter is the expectation value of the traced Polyakov
loop, $\lLA$, that is related to the free energy of a
quark--anti-quark pair. It has been computed on the lattice for both,
Yang-Mills theory and full QCD at finite temperature, but quantitative
continuum computations had been missing so far. A further order
parameter is related to the expectation value $\bar\varphi$ of the
algebra-valued field $\varphi$, built from the expectation value of
the gauge invariant eigenvalues, see the discussion around
\eq{eq:barvarphi}. With $\bar\varphi$ we can also compute the traced
Polyakov loop $L(\bar\varphi)$, an order parameter that bears
similarities to $\lLA$. The field $\varphi$ is directly related to the
gauge field and its expectation values are $\langle A_0\rangle$. It
has first been computed non-perturbatively within continuum approaches
to Yang-Mills theory and QCD in \cite{Braun:2007bx, Marhauser:2008fz,
  Braun:2009gm, Braun:2010cy, Fister:2013bh}, first lattice results
have been presented e.g.\ in
\cite{Langfeld:2013xbf,Diakonov:2013lja}. It has also been argued that
Polyakov-loop enhanced low-energy effective models should rather be
based on the latter observables, $\bar\varphi$ and $L(\bar\varphi)$,
than on the former, $\lLA$, see e.g.\ \cite{Braun:2009gm,
  Pawlowski:2010ht,Herbst:2013ail,Haas:2013qwp,Herbst:2013ufa,Pawlowski:2014aha}.

We have argued that the order parameters $\lLA$ and $L(\bar\varphi)$
agree within Gau\ss ian approximations for correlations of the
temporal gauge field $A_0$ or $\varphi$, and consequently within such
an approximation for the correlations of $L$, see the discussion around
Eqs.~\eq{eq:Jensen} to \eq{eq:Gausstr} on page 5. Low-energy effective 
models built on $\lLA$ rest on these Gau\ss ian approximations
for the computation of the quark-gauge field or rather quark-Polyakov
loop fluctuations. In contradistinction, models built on
$\bar\varphi$ are not subject to these Gau\ss ian approximations as
they can be derived directly from QCD. The difference between $\lLA$
and $L(\bar\varphi)$ is a direct measure for the non-Gau\ss ianity of
these fluctuations, and hence a measure for the quantitative
reliability of $\lLA$-based models.

In the present work we have discussed the differences and similarities
of these observables. We have also provided a review of the underlying
symmetries and in particular of the connection of the algebra-valued
field $\varphi$ to the gauge field, which is the basis of continuum
formulations of Yang-Mills theories. The background glue potential in
Yang-Mills theory has been computed within the functional
renormalisation group, similarly to
\cite{Braun:2007bx,Braun:2010cy,Fister:2013bh}. Using the flow
equation for general composite operators put forward in
\cite{Pawlowski:2005xe} we have furthermore derived a flow equation
for $\lLA$. This flow equation is fully non-perturbative and also
utilises the expectation value $\bar\varphi$. The difficult task of
computing infinite-order correlation functions of the gauge field is
resolved within the successive integration of momentum fluctuations:
each iterative infinitesimal momentum-shell integration increases the
order of gauge field fluctuations taken into account.

We have shown in \sec{sec:Renormalization} that the present
continuum computation of $\lLA$ is renormalisation group
invariant. Moreover, the present functional renormalisation group
approach provides a direct and simple access to the renormalisation
procedure. This also allows us to discuss the relative renormalisation
between the lattice and present continuum computations for $\lLA$. It
has been argued that it amounts to a temperature-dependent
renormalisation at large temperatures, see \sec{sec:Renormalization}.

In \sec{sec:Comparison} we have shown that the results for
$\lLA$ within the present non-perturbative continuum approach agree
quantitatively with the lattice results, see \fig{fig:PLlat_T}. As the
present continuum computation of $\lLA$ is based on the input
$\bar\varphi$ or rather $\lA0$, it also is a further non-trivial
support for the quantitative reliability of the results for $\lA0$ in
\cite{Braun:2007bx, Marhauser:2008fz, Braun:2009gm, Braun:2010cy,
  Fister:2013bh}, and related applications in QCD-enhanced low energy
effective models \cite{Herbst:2013ail,Haas:2013qwp,Herbst:2013ufa}.

We have shown that the non-Gau\ss ianity of the fluctuations grows
strong in the regime $T\lesssim 3 T_c$ which is the regime of interest
for the low energy effective models, see \fig{fig:PL_ren}. This
suggests that, in particular for an analysis of fluctuations in QCD,
it is important to take into account the Polyakov loop fluctuations if
one aims at quantitative precision. This can either be done by using
models based on $\bar\varphi$ or $\lA0$, or by amending the
$\lLA$-based effective models by a fluctuation analysis of the
Polyakov loop.

In turn, both order parameters are well described by (resummed)
perturbation theory for temperatures $T\gtrsim 3 T_c$. Such a
behaviour is already well-known from the trace anomaly. In the present
continuum approach in the Landau-deWitt gauge it can be traced back to
the non-perturbative mass-gap in the gluon propagator. Even though the
propagator is a gauge-dependent quantity, this mass-gap reflects the
mass-gap in Yang-Mills theory, and is directly related to confinement.

Finally, the extension of the presented approach to other gauge
groups, such as $SU(2)$ or $SU(N)$ with $N>3$ and exceptional Lie
groups, as well full QCD is straightforward and relies solely on the
knowledge of the corresponding propagators. Here, the case of QCD with
$N_f$ quark flavours is of particular interest: the corresponding
unquenched propagators and unquenching of the glue potential have been
put forward in \cite{Braun:2009gm, Fischer:2013eca}. The structure of
the flow equation \eq{eq:flowL} then suggests that a splitting between
$\lL$ and $\lLc$ will arise naturally, accounting for the fact that
these objects are related to the free energies of quarks and
anti-quarks, respectively. Furthermore, since light quarks break the
center symmetry explicitly, the confinement-deconfinement transition
is a crossover in this case. The temperature-dependent differences
between $L[\lA0]$ and $\lL$, as observed here, then have a drastic
impact on the pseudo-critical temperature, usually deduced from the
inflection point of these observables. Indeed, the order parameter
$L(\lA0)$ has been computed with continuum methods in $N_f=2$ flavour
QCD in \cite{Braun:2009gm} for finite temperature and imaginary
chemical potential, and in $N_f=2+1$ flavour QCD in
\cite{Fischer:2013eca} for finite temperature and real chemical
potential. As expected, both show a significantly stepper thermal rise
in comparison to the lattice results for $\lL$, see e.g.\
\cite{Borsanyi:2010bp} for $N_f =2+1$ flavour results.  Moreover, the
pseudo-critical temperature of $L[\lA0]$ for the
confinement-deconfinement phase transition agrees well with the chiral
pseudo-critical temperature, see \cite{Braun:2009gm,Fischer:2013eca}.
From the present work it is clear that this originates in the
different treatment of non-Gau\ss ian fluctuations in the order
parameters. The close chiral and confinement-deconfinement transition
temperatures also suggests that the observable $L[\lA0]$ shows a
closer resemblance to baryonic fluctuation observables that have a
direct physics interpretation in dynamical QCD. 

In summary, this work lays the foundation for a systematic study of
the confinement-deconfinement transition in the phase diagram of QCD,
both in model approaches and from first principles.

%%%%%%%%%%%%%%%%%%%%%%%%%%%%%%%%%%%%%%%%%%%%%%%%%%%%%
%%%%%%%%%%%%%%%%%%%%%%%%%%%%%%%%%%%%%%%%%%%%%%%
\subsection*{Acknowledgements}
We thank the fQCD collaboration, in particular L.~Fister, for
discussions and collaboration on related topics. This work is
supported by the Helmholtz Alliance HA216/EMMI, by ERC-AdG-290623 and
by the BMBF grant OSPL2VHCTG.

%%%%%%%%%%%%%%%%%%%%%%%%%%%%%%%%%%%%%%%%%%%%%%%%%%%%%%%%%%%%%%%%%%%%%%%%%%%%%%%%%%%%%%%%%%%%%%%%%%%%
% References
%%%%%%%%%%%%%%%%%%%%%%%%%%%%%%%%%%%%%%%%%%%%%%%%%%%%%%%%%%%%%%%%%%%%%%%%%%%%%%%%%%%%%%%%%%%%%%%%%%%%
\bibliography{../refs}

\end{document}